\documentclass[12pt]{article}

\usepackage{scicite}
\usepackage{times}

\usepackage{graphicx}
\usepackage{subfigure}
\usepackage{float}
\usepackage{url}
\usepackage{gensymb}
\usepackage{gensymb}

\newcounter{lastnote}

\title{The Weather Impacts the Outbreak of COVID-19 in Mainland China}

\author
{Siyu Huang, Ji Liu, Haoyi Xiong, Jizhou Huang, Haozhe An, Dejing Dou\thanks{Corresponding author}\\
\\
\normalsize{Baidu Inc., Beijing, China} \\ 
\\
\normalsize{
\{huangsiyu, liuji04, xionghaoyi, huangjizhou01, v\_anhaozhe, doudejing\}@baidu.com
}
}

\begin{document}

\maketitle


\begin{abstract}
Recent literature has suggested that climate conditions have considerably significant influences on the transmission of coronavirus COVID-19. However, there is a lack of comprehensive study that investigates the relationships between multiple weather factors and the development of COVID-19 pandemic while excluding the impact of social factors. In this paper, we study the relationships between six main weather factors and the infection statistics of COVID-19 on 250 cities in Mainland China. Our correlation analysis using weather and infection statistics indicates that all the studied weather factors are correlated with the spread of COVID-19, where precipitation shows the strongest correlation. We also build a weather-aware predictive model that forecasts the number of infected cases should there be a second wave of the outbreak in Mainland China. Our predicted results show that cities located in different geographical areas are likely to be challenged with the second wave of COVID-19 at very different time periods and the severity of the outbreak varies to a large degree, in correspondence with the varying weather conditions.
\end{abstract}

\section{Introduction}

Since December 2019, the coronavirus COVID-19 pandemic has rapidly spread all over the world. The number of confirmed cases of COVID-19 increases sub-exponentially in China \cite{Maier2020}, and some other countries. As of June 6, the total number of confirmed cases saturated at around 84.6 thousand in China with about 10 new cases per day while it still rapidly increases in other countries, e.g. Brazil, United States, and India. 

In Mainland China, it is generally assumed that many factors have influences on the spread of the COVID-19 pandemic, e.g. the containment measures \cite{Maier2020} and the inflows from Wuhan \cite{Xiong2020}. However, the impact of weather on COVID-19 at city scale is also interesting to analyze in order to reveal the spread of the COVID-19 pandemic. The correlation between the weather condition and the transmission of COVID-19 can help provide an accurate estimation, which is important information for the public citizens and the authorities. 

The relationships between climate conditions and the transmission of influenza have been extensively studied in a series of existing literature \cite{fuhrmann2010effects,shaman2010absolute,shaman2011absolute,willem2012nice,shoji2011absolute}. Separated relationship between individual climate factors and the spread of the COVID-19 pandemic is studied in many recent works \cite{ma2020effects,tosepu2020correlation,sajadi2020temperature,gupta2020effect,csahin2020impact,luo2020role,liu2020impact,poirier2020role,chen2020roles,zhu2020association,shi2020impact}. For instance, temperature or humidity may have a significant correlation with the number of confirmed cases \cite{bukhari2020will}, such that there would be more confirmed cases of COVID-19 in a colder and wetter environment. On the other hand, public containment measures, including social distancing and wearing of masks, should also be implemented to reduce the transmission of COVID-19 even in summer \cite{wang2020high}. Oliveiros et al. \cite{oliveiros2020role} exploited more meteorological variables including temperature, humidity, precipitation, and wind speed to study the COVID-19 transmission patterns in China before February 29, 2020. In the study, linear regression analysis is conducted to investigate the correlation between the meteorological variables and the rate of increase and doubling time of the confirmed cases of COVID-19.

In this work, we aim at comprehensively studying the relationships between various weather factors (i.e., temperature, humidity, precipitation, wind speed, air pressure, visibility) and the infection statistics of COVID-19 pandemic, while excluding the impact of main social factors (i.e., population and inflow population from Wuhan). Widely used websites, e.g., Baidu Migration \footnote{Baidu Migration - http://qianxi.baidu.com/}, search engines, e.g., Baidu \footnote{Baidu - https://www.baidu.com/} and climate data website, e.g., online climate data \footnote{Online Climate Data  - https://rp5.ru/}, can be exploited to gather the data related to the aforementioned factors.
Then, based on the correlation analysis, we build a weather-based prediction model to estimate the infection rates of all the provinces in Mainland China if there is a second outbreak of coronavirus. We employ a multivariate regression model to transform the weather statistics to the number of confirmed cases in each province. More specifically, we would like to investigate the following problems: 

\begin{itemize}

    \item \textbf{To what degree do the local climate factors affect the pandemic outbreaks of COVID-19?} The correlations among individual weather factors, social factors, and the spread of the COVID-19 pandemic have been studied in existing literature \cite{al2020correlation, Xiong2020, Maier2020}. However, the correlations between the local climate factors and the outbreaks of the COVID-19 pandemic have not been comprehensively studied at a large scale up to six different weather factors and 250 cities. 
    
    \item \textbf{How to build a weather-aware prediction model for the second outbreak of COVID-19?} Based on the correlations between weather conditions and COVID-19 infection statistics, we propose to build a weather-aware prediction model for estimating the number of confirmed cases, if there is a second outbreak of COVID-19 in Mainland China. 
    
\end{itemize}

Different from the existing works, this paper aims to conduct a comprehensive study on the correlation among six different main weather factors and the confirmed cases of COVID-19 in 250 Chinese cities. For a deeper study, this paper incorporates the statistics of city populations and the inflow from Wuhan as controllable variables, so as to achieve a more accurate and convincing result on the influence of weather conditions on COVID-19 pandemic. The time-series analysis of precipitation and COVID-19 is also presented in the paper. Based on the correlation analysis, we project the number of confirmed cases at the province scale when there is a second outbreak of the COVID-19 pandemic.

\section{Method}

In this section, we present the method to carry out correlation analysis and the prediction method of coronavirus outbreak.

\subsection{Correlation Analysis}

In order to analyze the correlation between two random variables, we calculate the Pearson correlation coefficients \cite{Benesty2009} and conduct the Student's T-test (two tails) to verify the significance test (the same for the following analysis in the paper). 

We additionally employ partial correlation analysis \cite{Baba2004,Kenett2015} to obviate the impact of the population scale, i.e., the city population and the inflow population from Wuhan. In order to estimate the partial correlation of random variables $X$ and $Y$ with the random variable $Z$ removed, we compute the partial correlation coefficient in terms of the Pearson correlation coefficients as
\begin{equation}
    \rho (X,Y|Z) \equiv \frac{\rho (X,Y)-\rho (X,Z)\rho (Y,Z)}{\sqrt{(1-\rho ^2(X,Z))(1-\rho ^2(Y,Z))}} 
\end{equation}

\subsection{Prediction Method of Coronavirus Outbreak}
\label{sec:multivariate}
In this paper, we propose to predict the infection rate of the second outbreak of coronavirus using a multivariate regression model. The independent variables of the regression model are the statistics of six weather factors averaged from December 1, 2019 to February 15, 2020. The dependent variable is the accumulated COVID-19 infection rate before February 29, 2020. To alleviate the data randomness, we learn the regression model on the province-level data by aggregating the city-level data. In inference phase, the independent variables are the weather statistics averaged over the same one-month data in the history. The output of regression model denotes the accumulated infection rate until the end of next month. 

\section{Datasets}

In this section, we present the provenance of our datasets.

\subsection{COVID-19 Cases and Populations in 250 Chinese Cities}
\begin{figure}
    \subfigure[Accumulative confirmed cases of COVID-19]{ 
    \begin{minipage}[]{0.5\linewidth}
    \flushleft
    \includegraphics[width=0.96\linewidth]{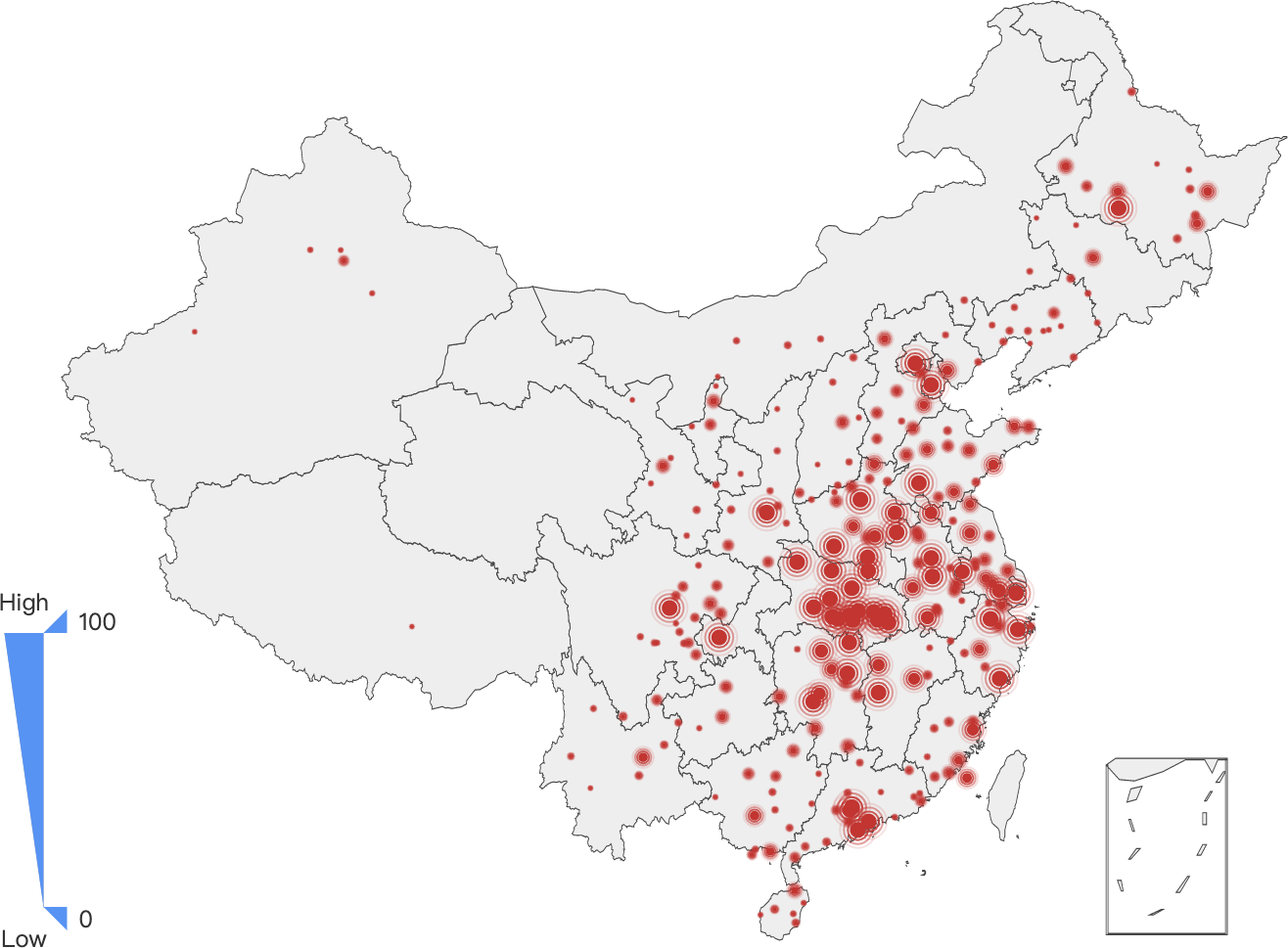}
    \vspace{1em}
    \end{minipage}
    }
    \subfigure[Inflow from Wuhan]{ 
    \begin{minipage}[]{0.5\linewidth}
    \flushright
    \includegraphics[width=0.96\linewidth]{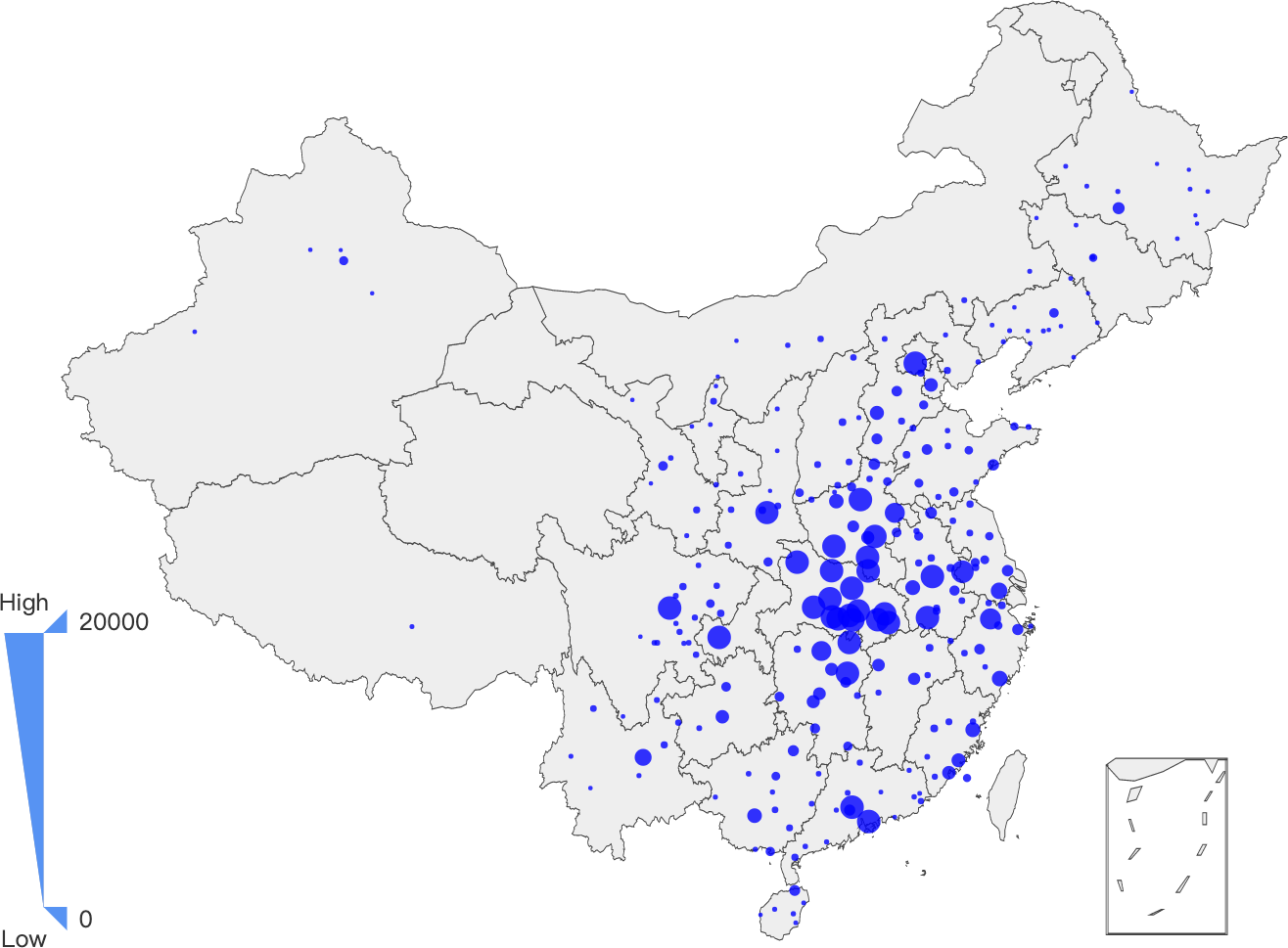}
    \vspace{1em}
    \end{minipage}
    }
    \caption{(a) Accumulative confirmed cases of COVID-19  until May 15 in Mainland China. (b) Inflow from Wuhan to 250 Chinese cities.}
    \label{fig:infection}
\end{figure}

Following the previous literature \cite{Xiong2020}, we collected the number of confirmed COVID-19 infected cases before Mar 31, 2020, for 250 Chinese cities except Wuhan from Baidu COVID-19 Page, and, the populations of those cities from Baidu Encyclopedia. The collected data of infection numbers is visualized in Fig. \ref{fig:infection}(a).

\subsection{Inter-City Mobility}
We collected the inter-city transitions between major Chinese cities during the period from 23 Jan 2020 (Wuhan lockdown) to 31 Mar 2020. We used anonymized localization request logs of Baidu Maps and detected inter-city mobility as the location change in two consequent localization requests. The transitions of inter-city mobility were categorized by the origin and destination (OD) pairs. The collected data of inflow from Wuhan is visualized in Fig. \ref{fig:infection}(b).

\subsection{Weather Statistics}
\begin{figure}
    \subfigure[Temperature (\degree C)]{ 
    \begin{minipage}[]{0.5\linewidth}
    \flushleft
    \includegraphics[width=0.96\linewidth]{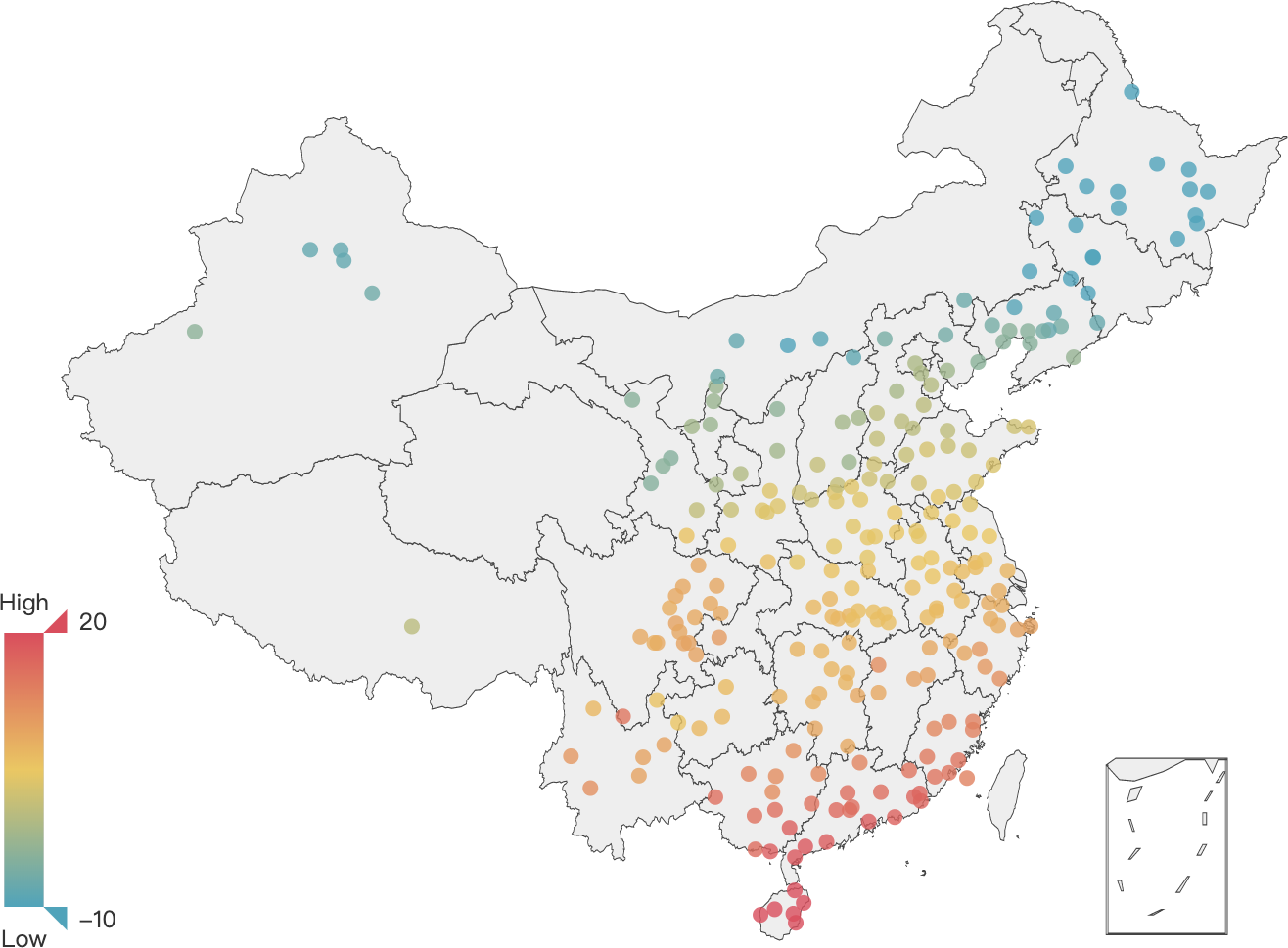}
    \vspace{1em}
    \end{minipage}
    }
    \subfigure[Relative Humidity (\%)]{ 
    \begin{minipage}[]{0.5\linewidth}
    \flushright
    \includegraphics[width=0.96\linewidth]{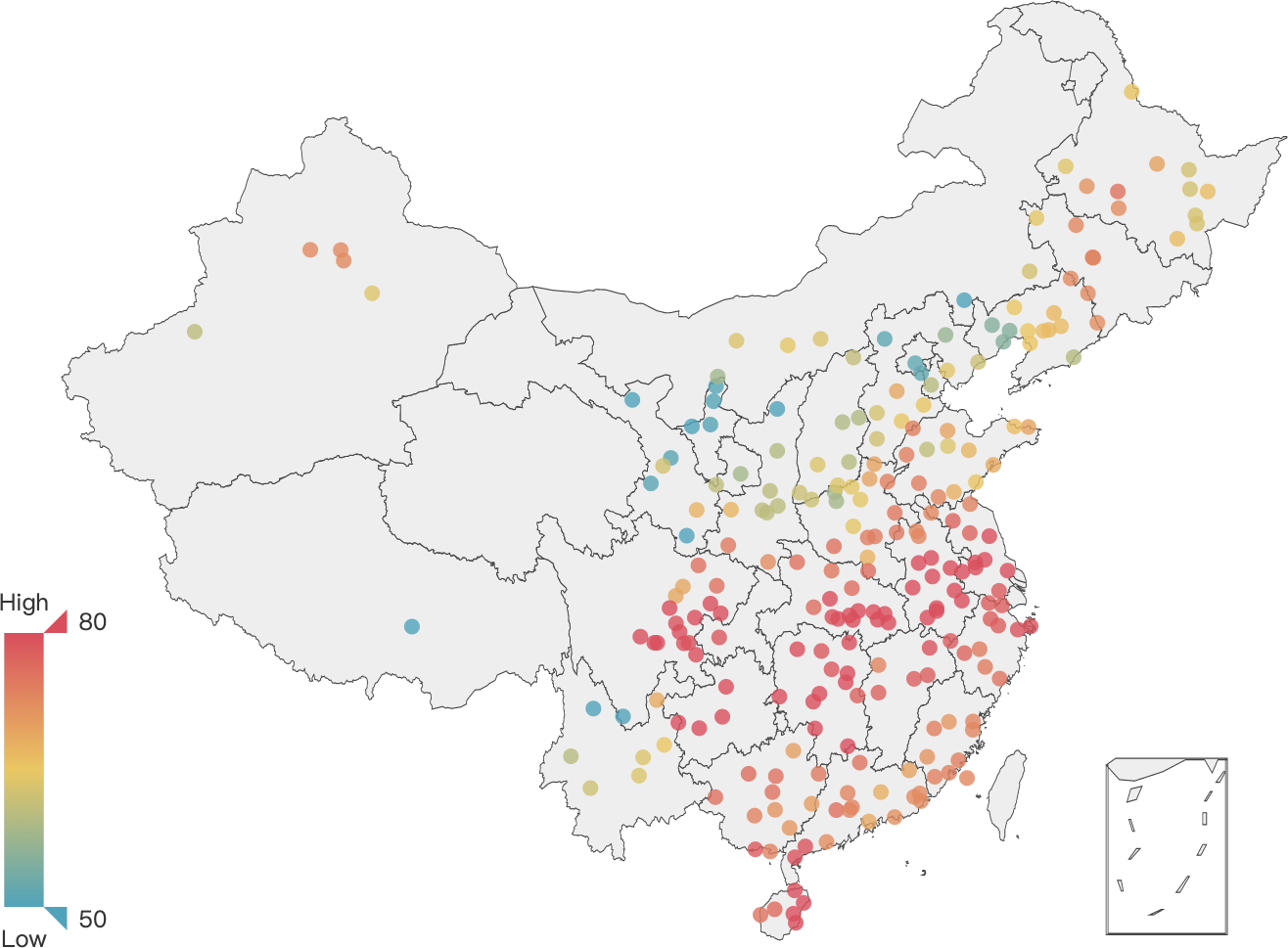}
    \vspace{1em}
    \end{minipage}
    }
    \\
    \vspace{0.5em}
    \\
    \subfigure[Precipitation (mm)]{ 
    \begin{minipage}[]{0.5\linewidth}
    \flushleft
    \includegraphics[width=0.96\linewidth]{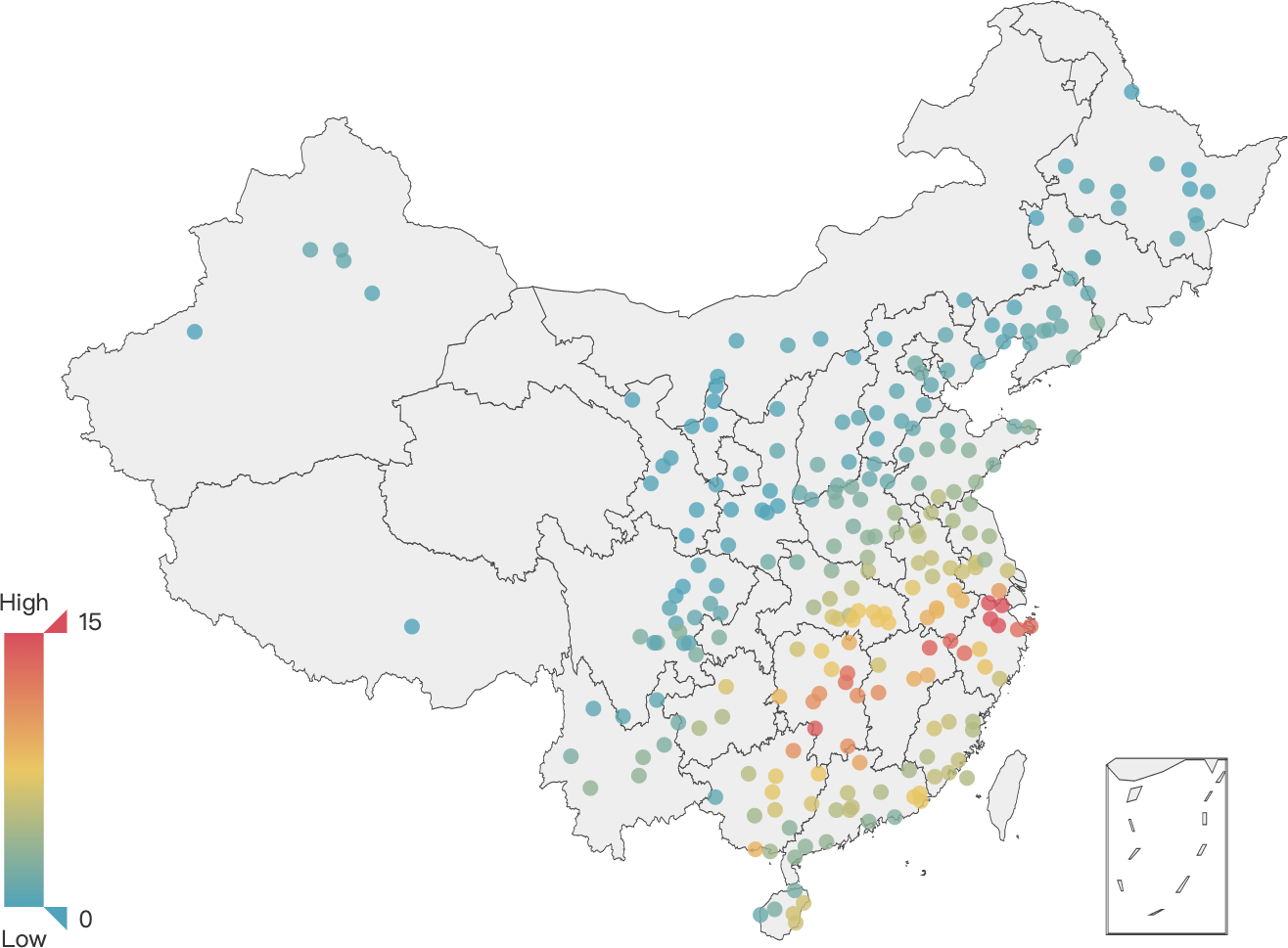}
    \vspace{1em}
    \end{minipage}
    }
    \subfigure[Wind Speed (m/s)]{ 
    \begin{minipage}[]{0.5\linewidth}
    \flushright
    \includegraphics[width=0.96\linewidth]{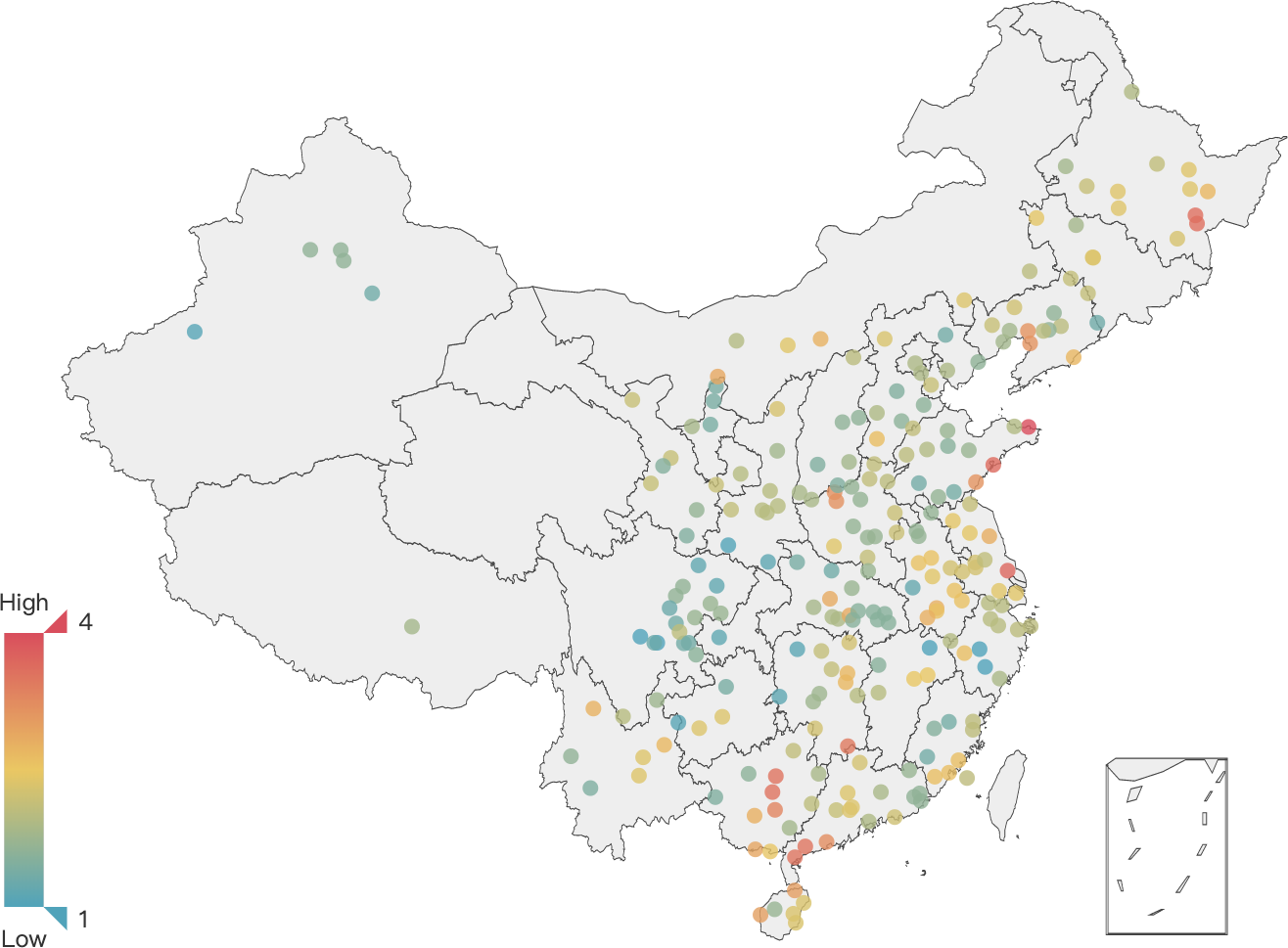}
    \vspace{1em}
    \end{minipage}
    }
    \subfigure[Air Pressure (mm)]{ 
    \begin{minipage}[]{0.5\linewidth}
    \flushright
    \includegraphics[width=0.96\linewidth]{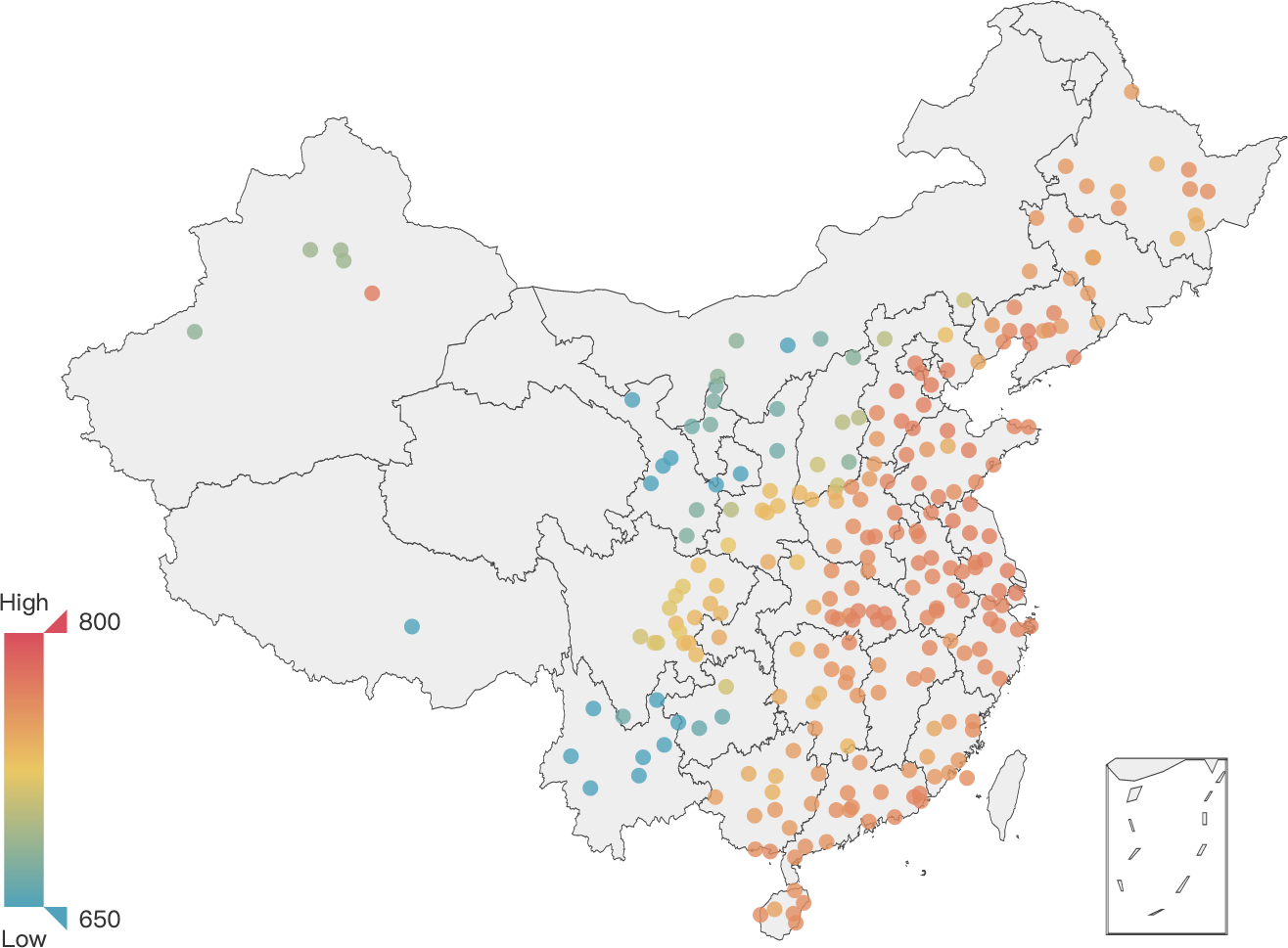}
    \vspace{1em}
    \end{minipage}
    }
    \subfigure[Visibility (km)]{ 
    \begin{minipage}[]{0.5\linewidth}
    \flushright
    \includegraphics[width=0.96\linewidth]{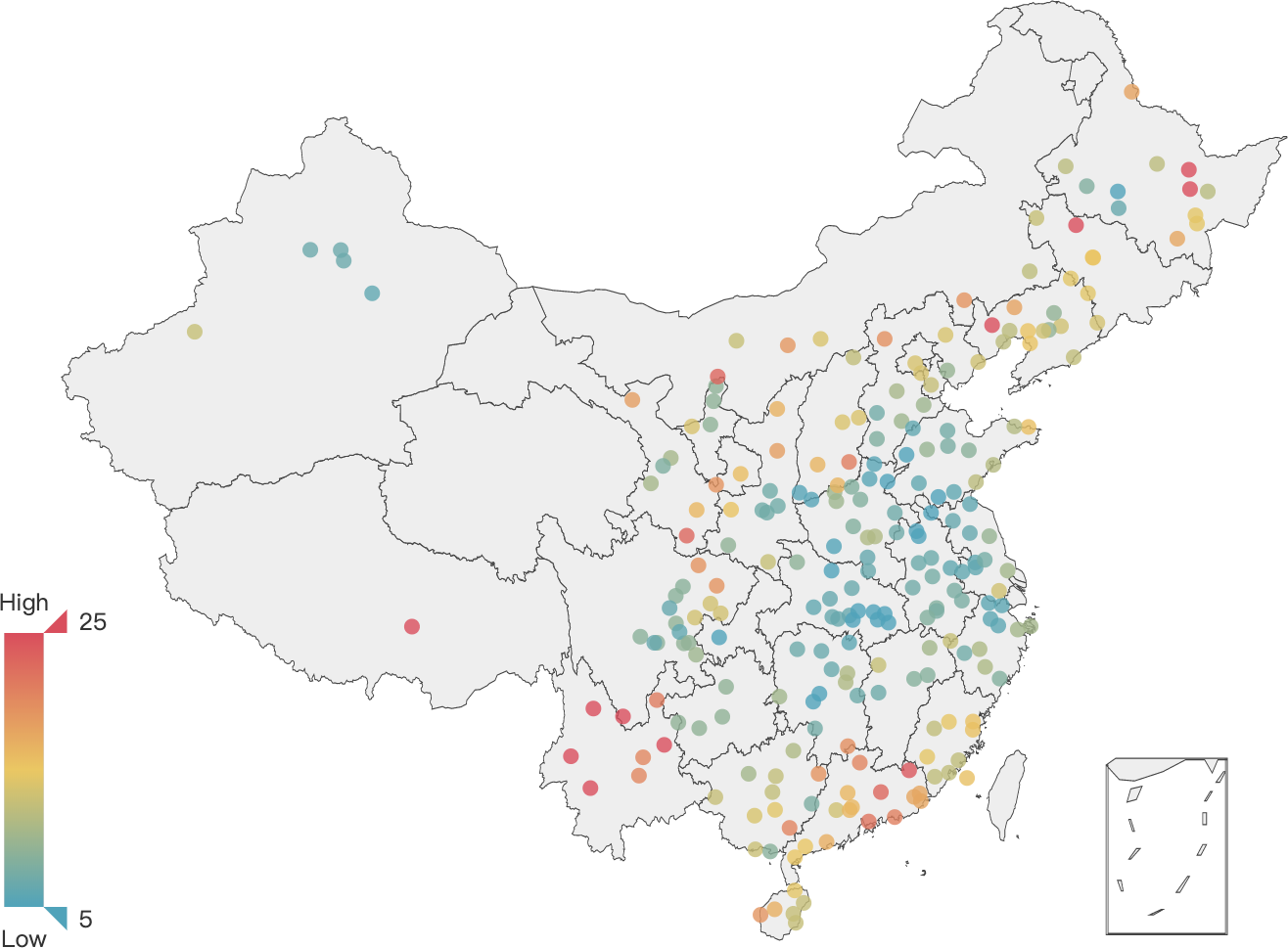}
    \vspace{1em}
    \end{minipage}
    }
    
    \caption{The statistics of six main weather factors of 250 Chinese cities, averaged from December 2019 to February 2020.}
    \label{fig:weather}
\end{figure}

We collected the weather statistics data from \url{www.rp5.ru}, which is a website providing weather observation data reported from more than 10,000 weather stations all over the world. We collected the weather data of 250 Chinese cities, from May 14 of 2010 to May 14 of 2020. The collected six weather factors are temperature, relative humidity, precipitation, wind speed, air pressure, and visibility. The collected weather data is visualized in Fig. \ref{fig:weather}.

\section{Results}

In this section we present the results observed from our study. We first show the correlations between different weather factors and the COVID-19 spread in Section \ref{sec:result_correlation}. We then present the time-series analysis on the relationships between precipitation and confirmed cases of COVID-19 in Section \ref{sec:result_precipitation}. At last, we present our estimation results on the second outbreak of COVID-19 in Mainland China in Section \ref{sec:result_prediction}, using our weather-based prediction model.

\subsection{Correlation Between Weather Conditions and COVID-19}
\label{sec:result_correlation}

\begin{figure}
    \centering
    \subfigure[Negative logarithm of $p$-value]{ 
    \includegraphics[width=1\textwidth]{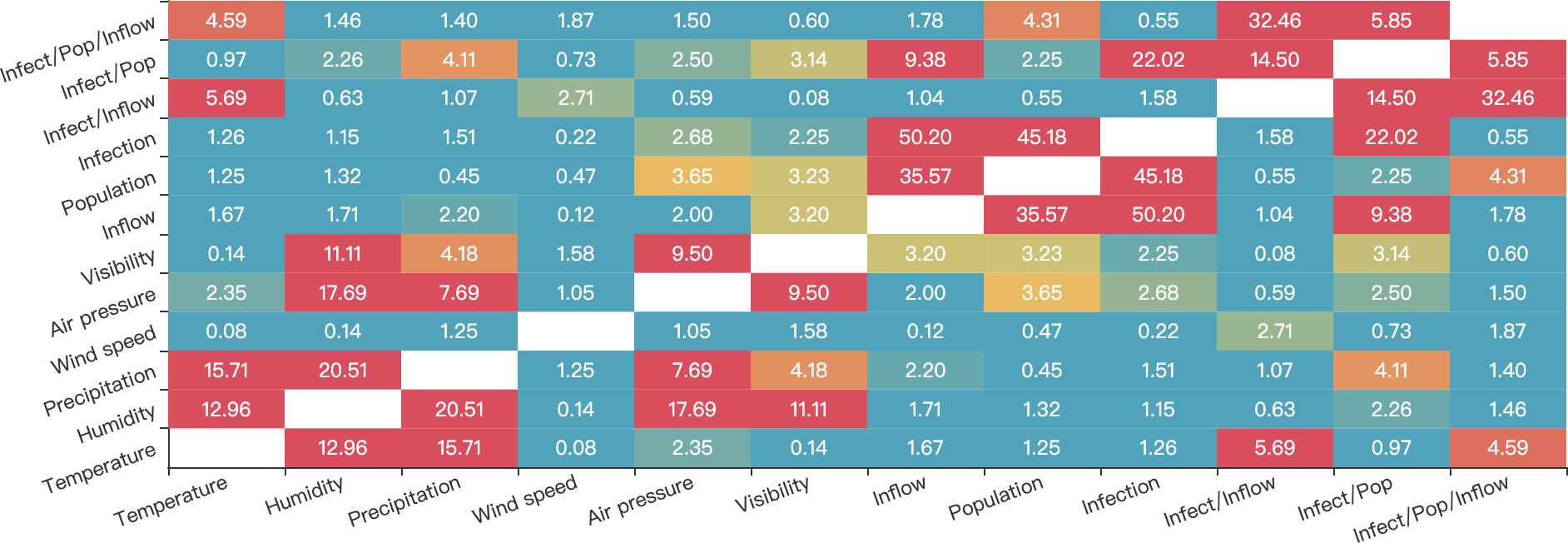}
    }
    \\
    \vspace{2em}
    \subfigure[Pearson correlation (R)]{ 
    \includegraphics[width=1\textwidth]{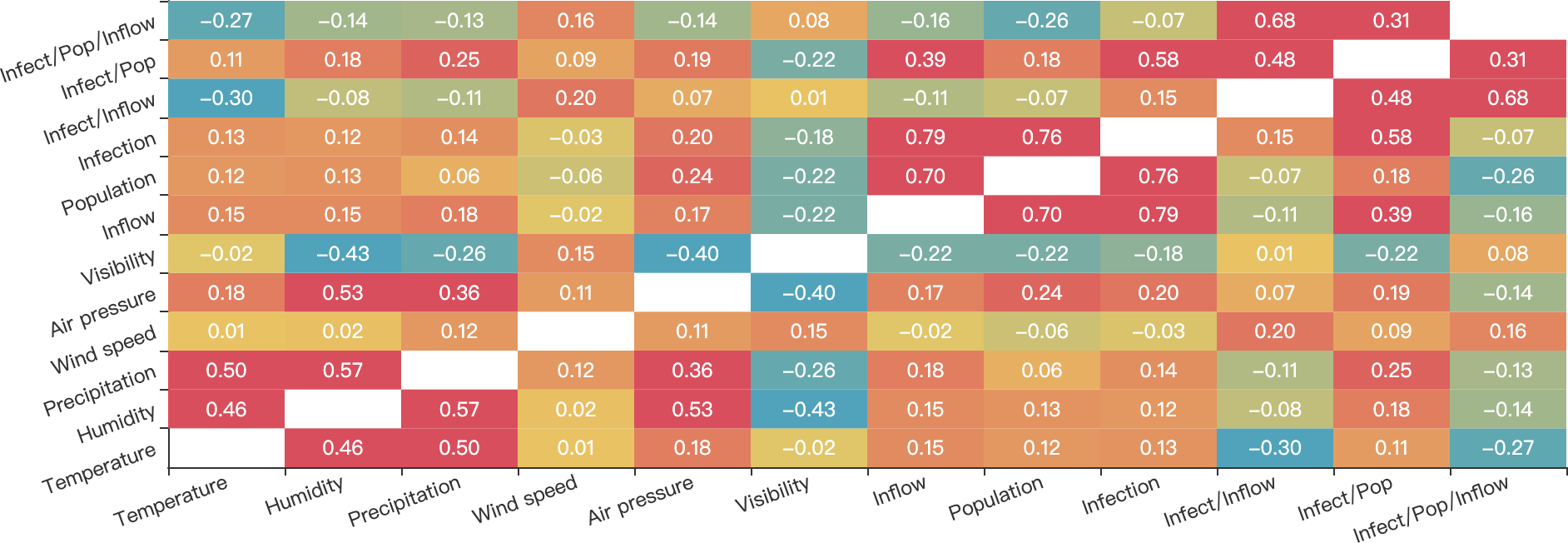}
    }
    \caption{(a) Negative logarithm of $p$-value and (b) Pearson correlation (R) between weather factors, social factors, and COVID-19 statistics.}
    \label{fig:pvalue}
\end{figure}

\emph{We compute correlations between the six weather factors and infection statistics of COVID-19. Results suggest that various weather factors have correlations with the infection statistics.}
We comprehensively investigate the correlation between the six main weather factors and COVID-19 statistics at 250 Chinese cities, and the resulted $p$-values and Pearson correlations are shown in Fig. \ref{fig:pvalue}(a) and (b), respectively. For the benefit of visualization, in Fig. \ref{fig:pvalue}(a) we show the \textit{negative logarithm of $p$-values}, i.e., a larger number denotes a larger correlation between factors, and, the number larger than $2.00$ ($p$-value $<0.01$) denotes that the correlation of the paired factors is significant in statistic. In Fig. \ref{fig:pvalue}, we observe that various different weather factors are correlated with the infection statistics: temperature - local infection rate ($\frac{\mbox{infection number}}{\mbox{inflow number}}$, i.e., confirmed cases per immigrant from Wuhan), humidity - infection rate ($\frac{\mbox{infection number}}{\mbox{population number}}$), precipitation - infection rate, wind speed - local infection rate, air pressure - infection rate, visibility - infection rate. 

\begin{figure}
    \subfigure[Temperature - Local Infection]{ 
    \begin{minipage}[]{0.31\linewidth}
    \includegraphics[width=0.98\linewidth]{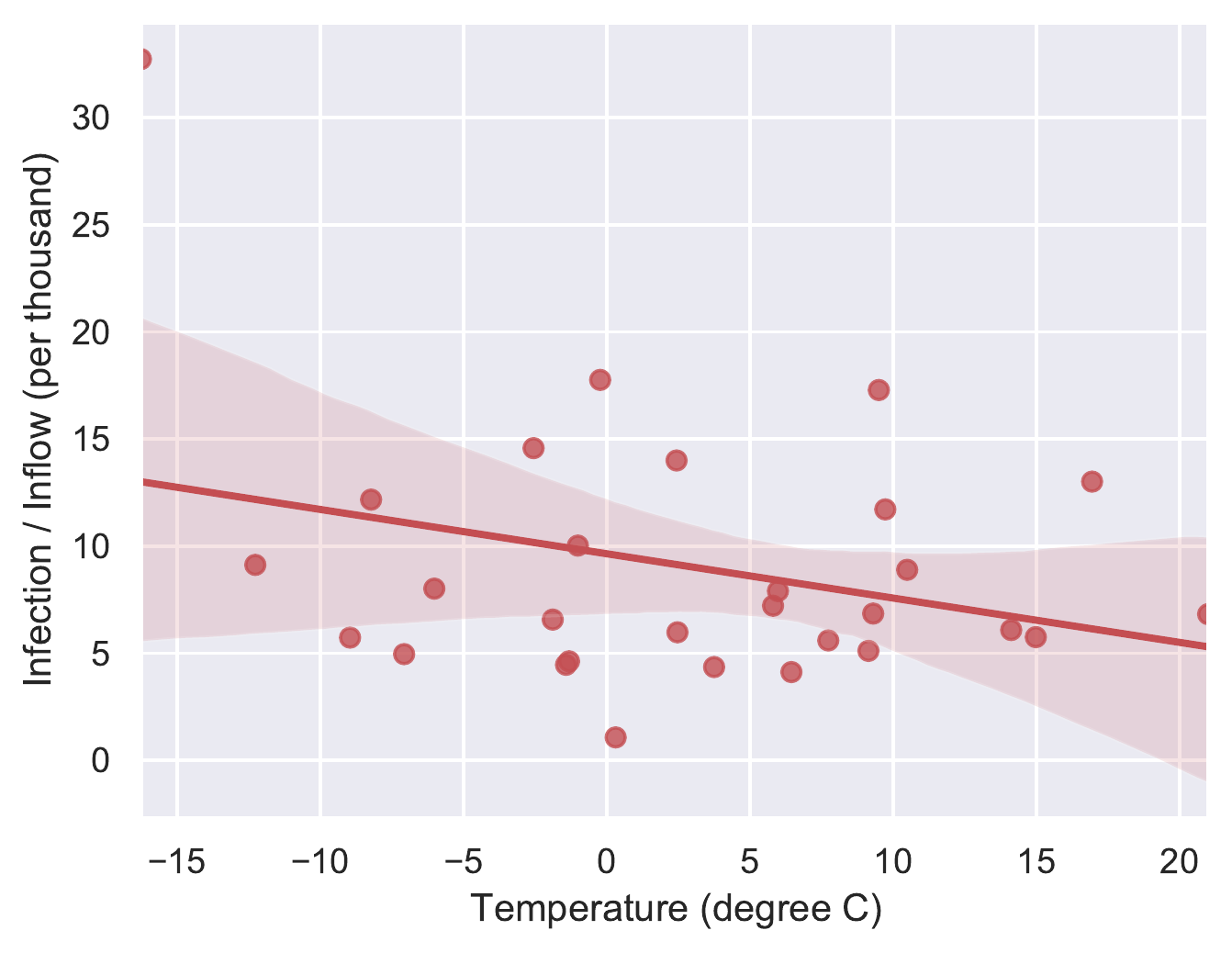}
    \vspace{0.5em}
    \end{minipage}
    }
    \subfigure[Humidity - Infection rate]{ 
    \begin{minipage}[]{0.31\linewidth}
    \includegraphics[width=0.98\linewidth]{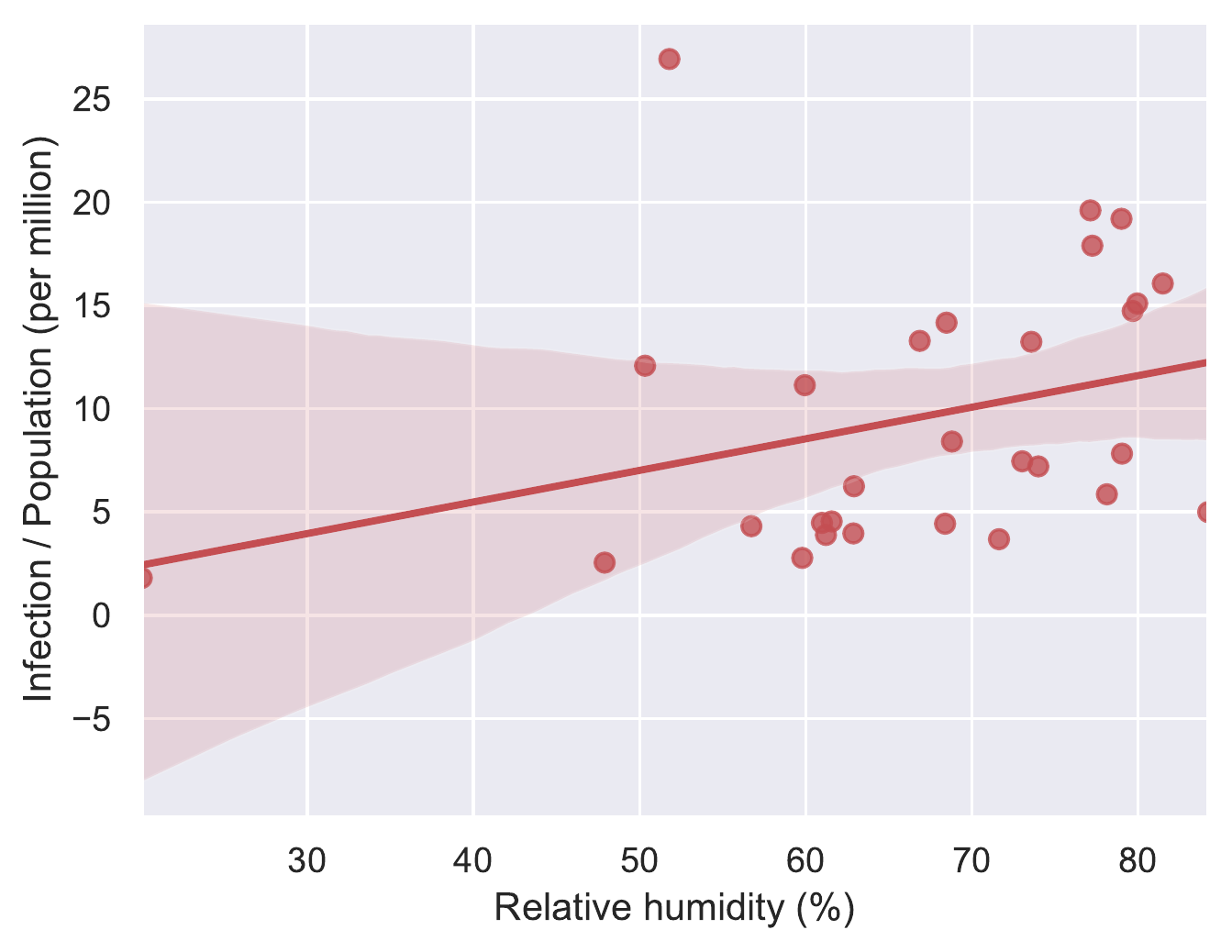}
    \vspace{0.5em}
    \end{minipage}
    }
    \subfigure[Precipitation - Infection rate]{ 
    \begin{minipage}[]{0.31\linewidth}
    \includegraphics[width=0.98\linewidth]{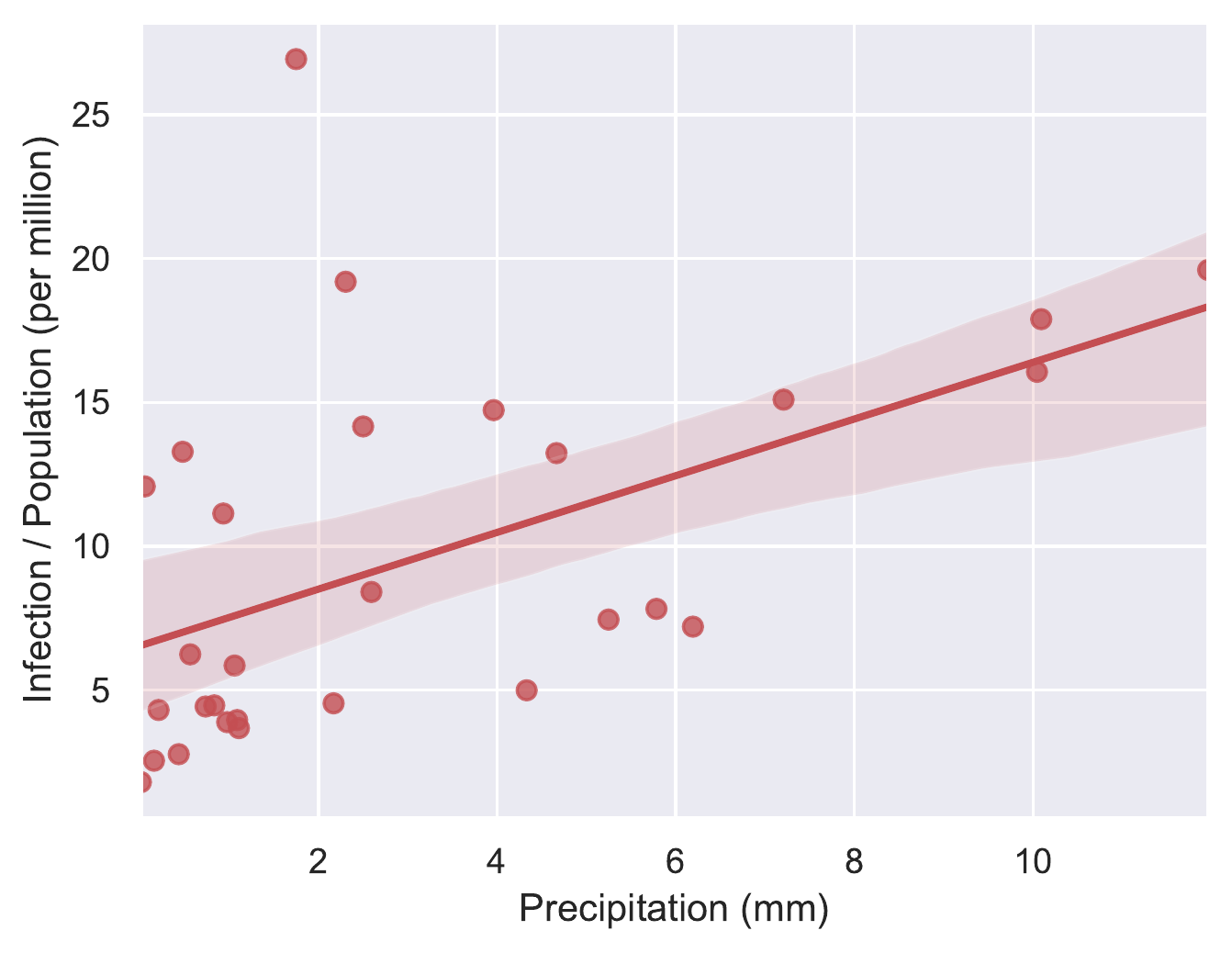}
    \vspace{0.5em}
    \end{minipage}
    }
    \\
    \vspace{0.3em}
    \\
    \subfigure[Wind speed - Local infection]{ 
    \begin{minipage}[]{0.31\linewidth}
    \includegraphics[width=0.98\linewidth]{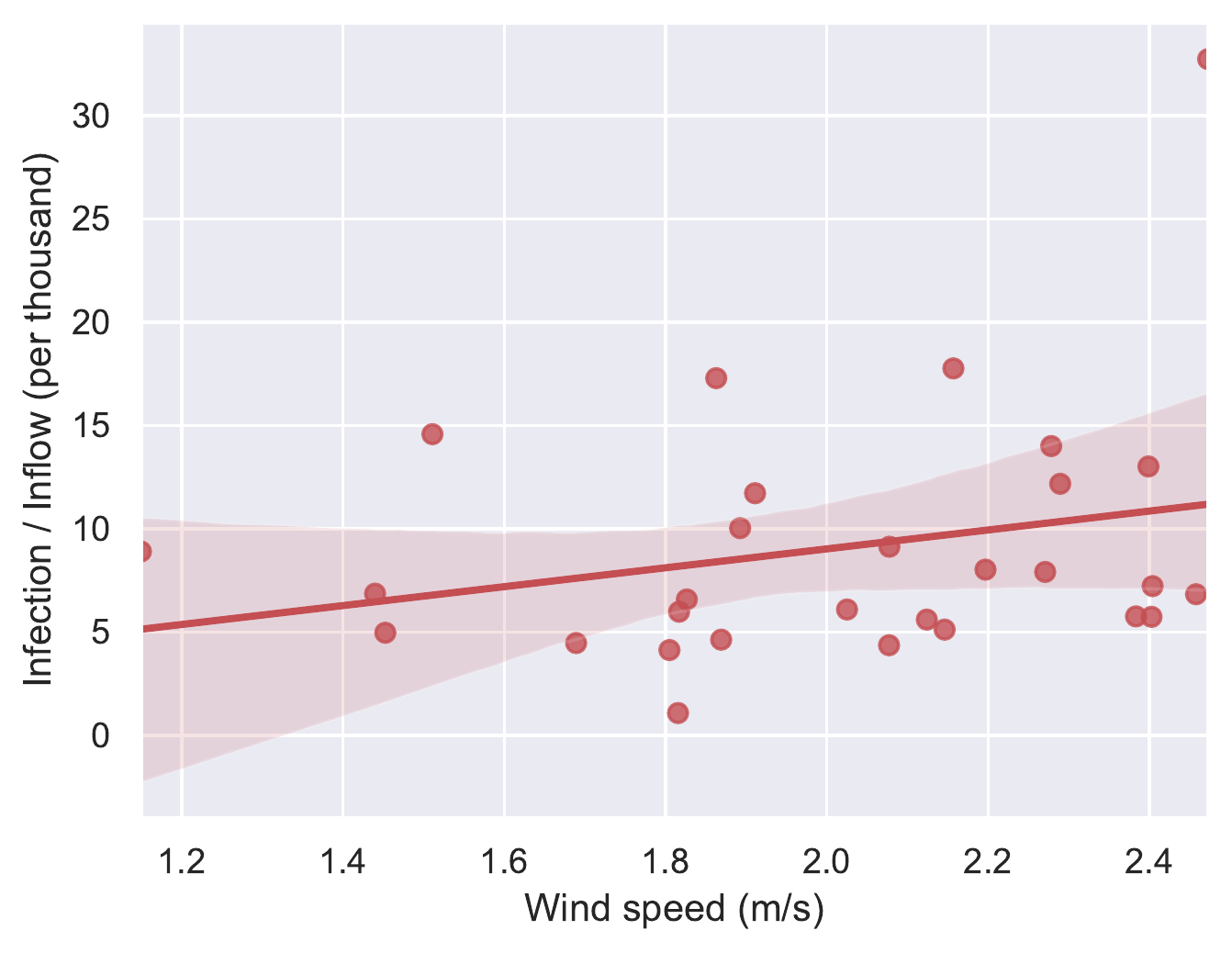}
    \vspace{0.5em}
    \end{minipage}
    }
    \subfigure[Air pressure - Infection rate]{ 
    \begin{minipage}[]{0.31\linewidth}
    \includegraphics[width=0.98\linewidth]{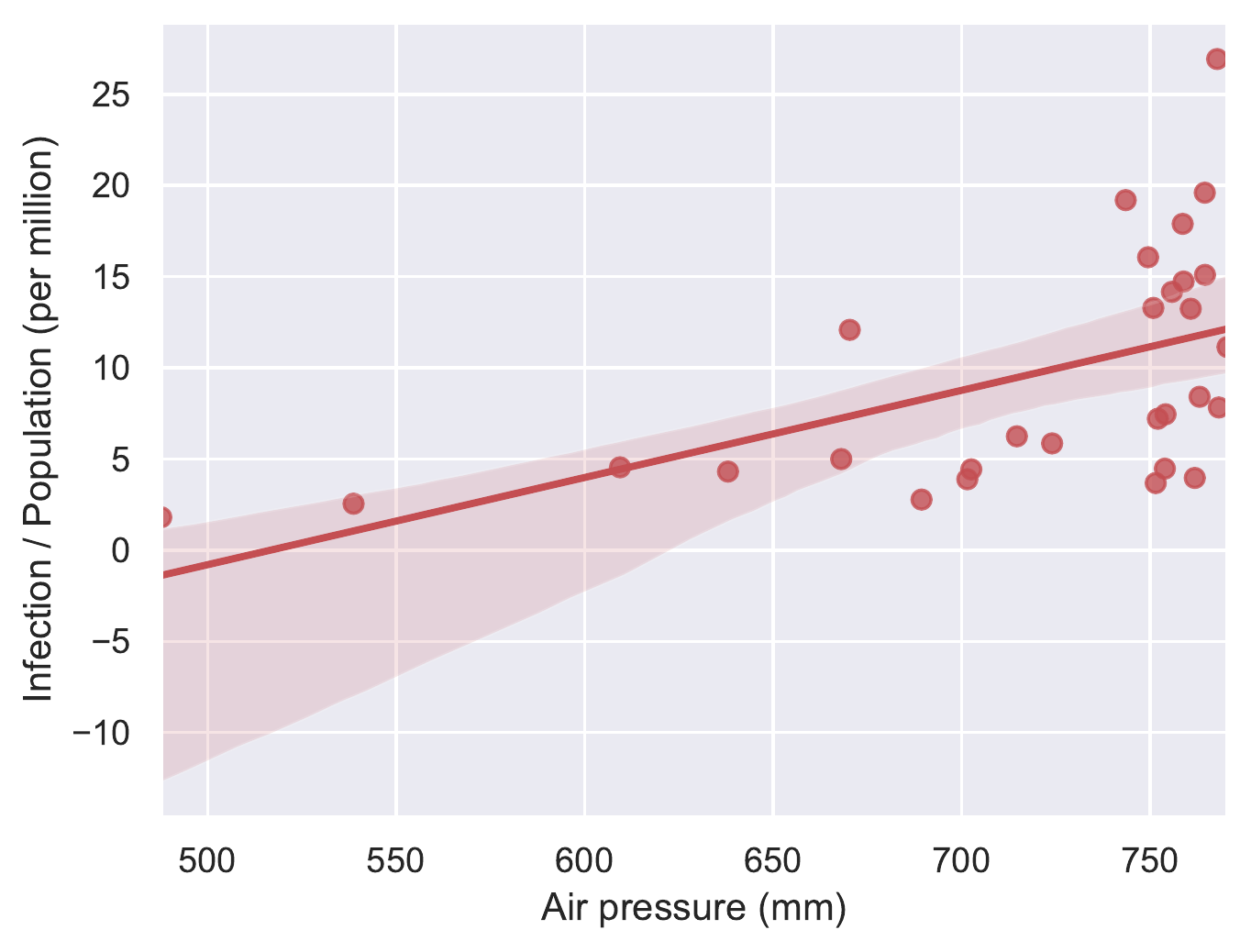}
    \vspace{0.5em}
    \end{minipage}
    }
    \subfigure[Visibility - Infection rate]{ 
    \begin{minipage}[]{0.31\linewidth}
    \includegraphics[width=0.98\linewidth]{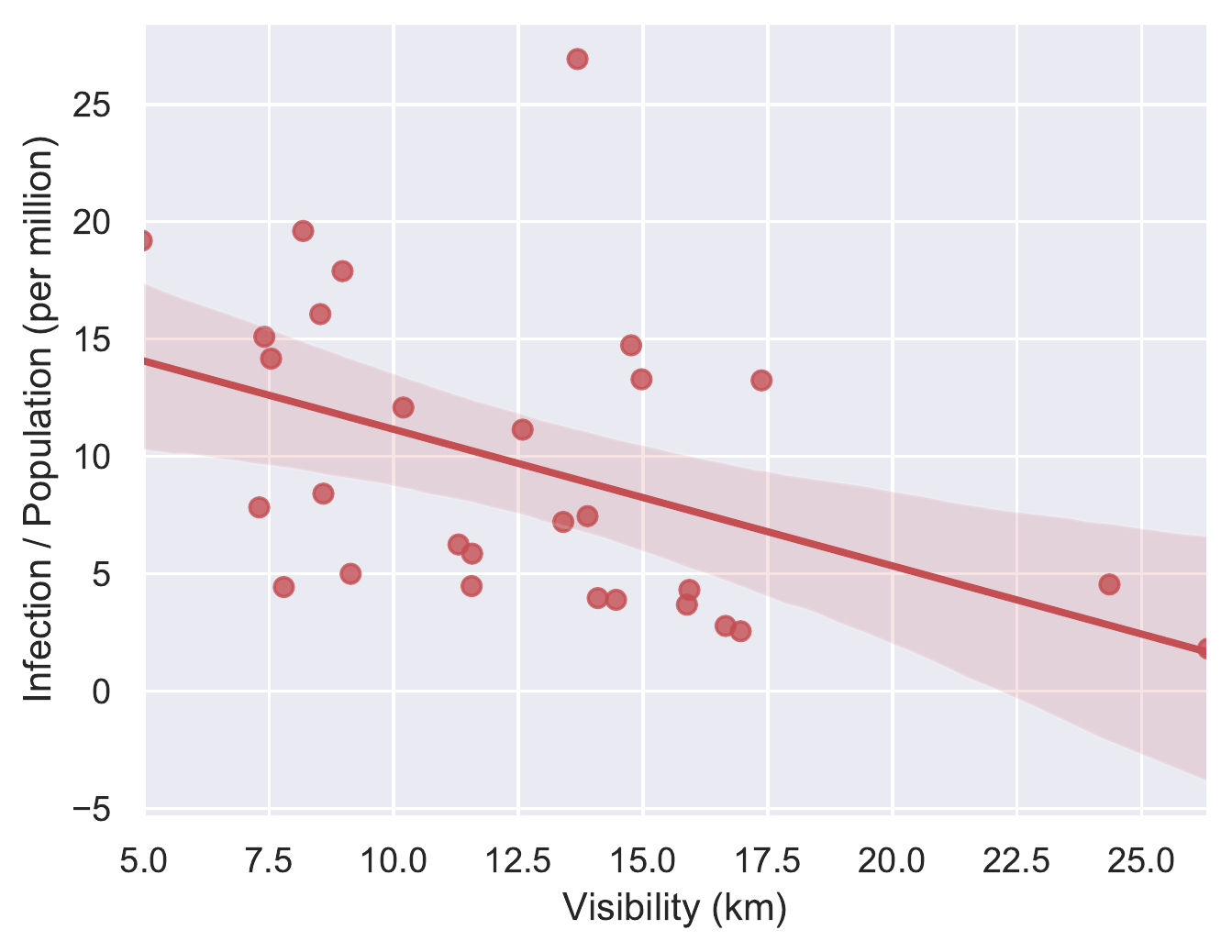}
    \vspace{0.5em}
    \end{minipage}
    }
    \caption{Linear regression results on weather factors and infection statistics.}
    \label{fig:confidence}
\end{figure}

Fig. \ref{fig:pvalue} only reveals the significance of correlation between the weather, social, and infection factors, but how do the weather and social factors affect COVID-19 pandemic? We show the linear regression results with 95\% confidence intervals in Fig. \ref{fig:confidence}. To alleviate the data randomness, this study is conducted on province-level by aggregating the city-level data. All six weather factors show clear regression slopes. The lower temperature and larger relative humidity induce the increase of local infection rate. The larger precipitation, higher wind speed, higher air pressure, and lower visibility induce the increase of infection rate.  

\emph{Precipitation is the most significant weather factor in the spread of COVID-19 in Mainland China.}
In Fig. \ref{fig:confidence}, the 95\% confidence intervals of temperature, humidity, and wind speed are not consistent among the samples. We conclude that only precipitation, air pressure, and visibility are consistently correlated with the COVID-19 statistics. Further, the air pressure and visibility are correlated with the population as denoted in Fig. \ref{fig:pvalue}. In conclusion, precipitation may be the most significant weather factor in the transmission of COVID-19.

\begin{figure}
    \subfigure[Population - Inflow]{ 
    \begin{minipage}[]{0.31\linewidth}
    \includegraphics[width=0.98\linewidth]{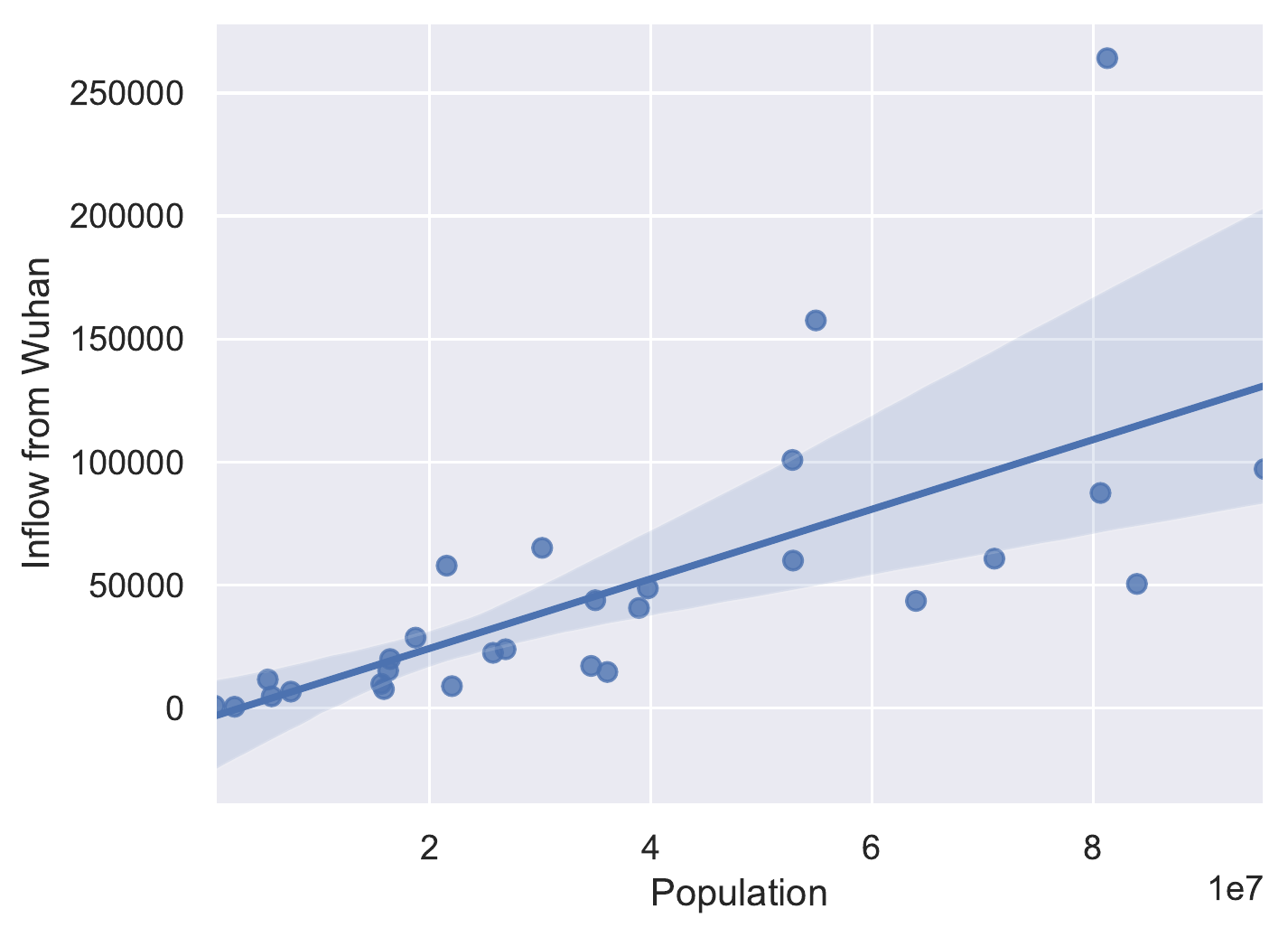}
    \vspace{0.5em}
    \end{minipage}
    }
    \subfigure[Population - Infection]{ 
    \begin{minipage}[]{0.31\linewidth}
    \includegraphics[width=0.98\linewidth]{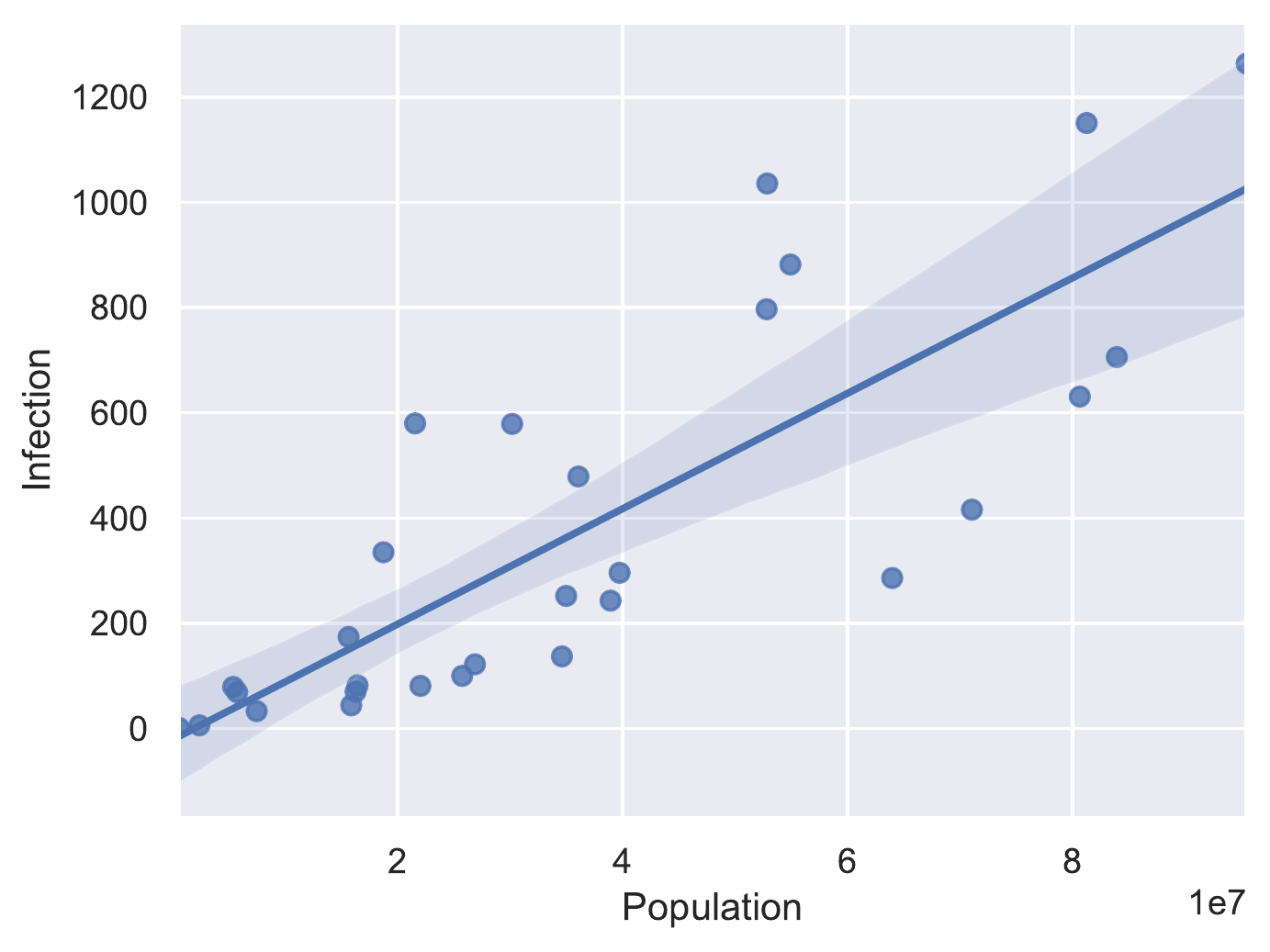}
    \vspace{0.5em}
    \end{minipage}
    }
    \subfigure[Inflow - Infection]{ 
    \begin{minipage}[]{0.31\linewidth}
    \includegraphics[width=0.98\linewidth]{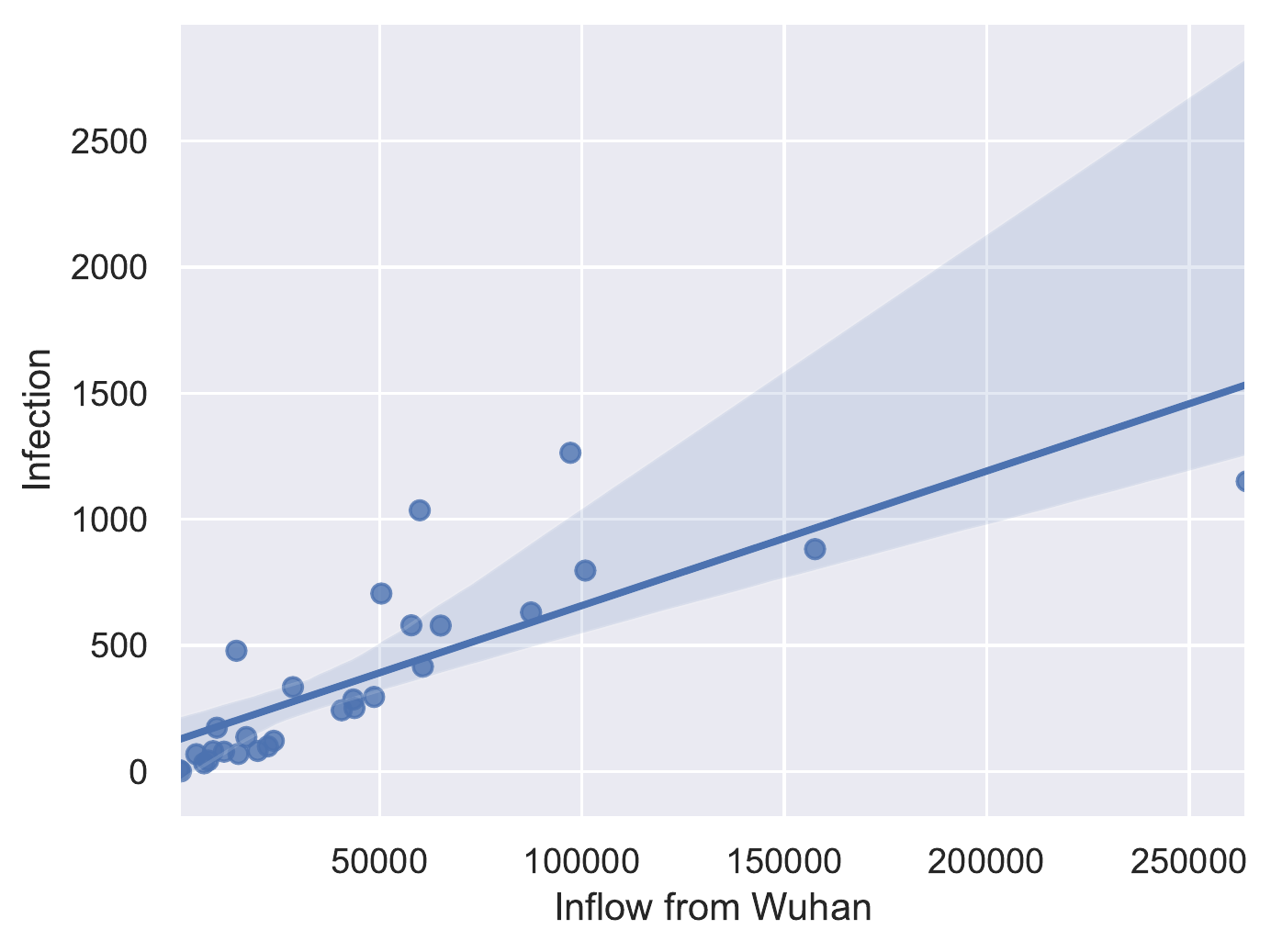}
    \vspace{0.5em}
    \end{minipage}
    }
    \caption{Linear regression results on social factors and confirmed infection cases.}
    \label{fig:confidence_social}
\end{figure}

\emph{The confirmed case of COVID-19 in Mainland China is highly correlated with the inflow population from Wuhan.}
We observe that there are strong correlations between the social factors and the the infection statistics. As shown in Fig. \ref{fig:confidence_social}(c) and the first part of Table \ref{tab:partial}, there is a very strong correlation between inflow from Wuhan and the infection number (R = $0.79$, $p$-value = $6.5\times10^{-51}$). Even when all six weather factors are treated as control variables, the inflow from Wuhan also appears to have a strong correlation with the infection rate (R= $0.35$, $p$-value= $4.8\times10^{-8}$). 

\emph{We validate the significance of precipitation in COVID-19 pandemic through partial correlation analysis.}
We conduct partial correlation analysis to study the impact of weather factors on infection rate by excluding the factor of inflow from Wuhan. The precipitation still shows a significant correlation with the infection rate (R=$20.5\%$, $p$-value=$1.6\times10^{-3}$). The fact reveals that precipitation is the most significant weather factor in the COVID-19 pandemic in China.

\subsection{Time-series analysis on precipitation in China, 2020}
\label{sec:result_precipitation}

\begin{figure}
    \subfigure[Historical Precipitation of Wuhan]{ 
    \begin{minipage}[]{0.5\linewidth}
    \flushleft
    \includegraphics[width=0.98\linewidth]{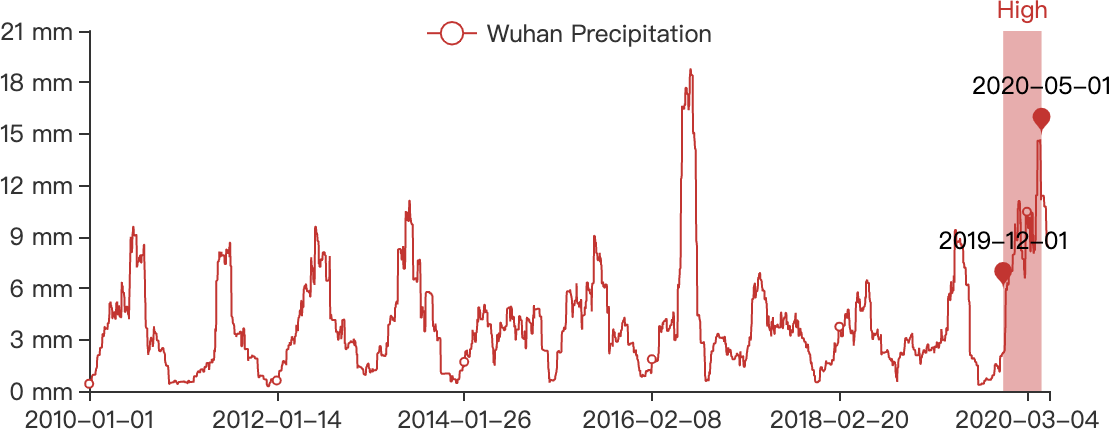}
    \vspace{0.3em}
    \end{minipage}
    }
    \subfigure[Historical Precipitation of China]{ 
    \begin{minipage}[]{0.5\linewidth}
    \flushright
    \includegraphics[width=0.98\linewidth]{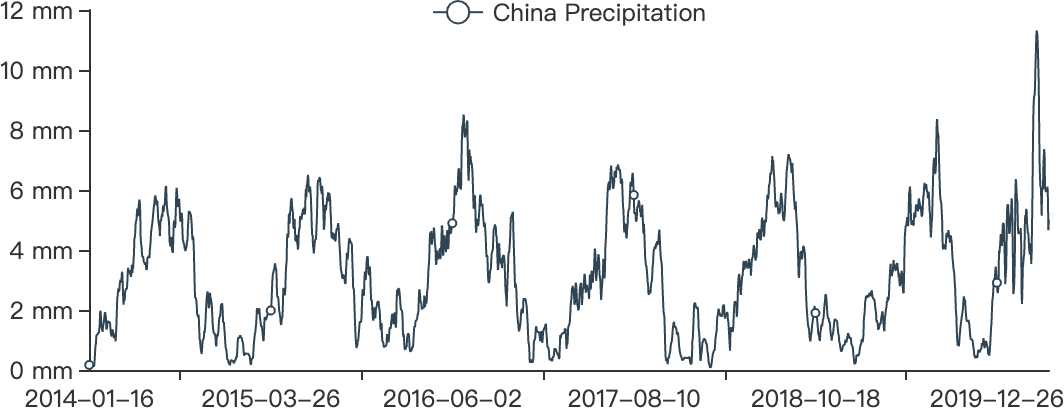}
    \vspace{0.3em}
    \end{minipage}
    }
    \\
    \vspace{0.3em}
    \\
    \subfigure[Precipitation in 2020 and History, Wuhan]{ 
    \begin{minipage}[ ]{0.5\linewidth}
    \flushleft
    \includegraphics[width=0.96\linewidth]{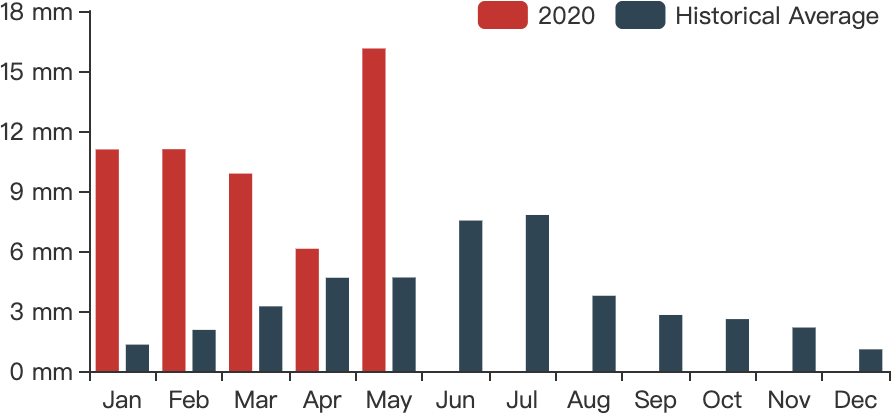}
    \vspace{0.3em}
    \end{minipage}
    }
    \subfigure[Precipitation in 2020 and History, China]{ 
    \begin{minipage}[]{0.5\linewidth}
    \flushright
    \includegraphics[width=0.96\linewidth]{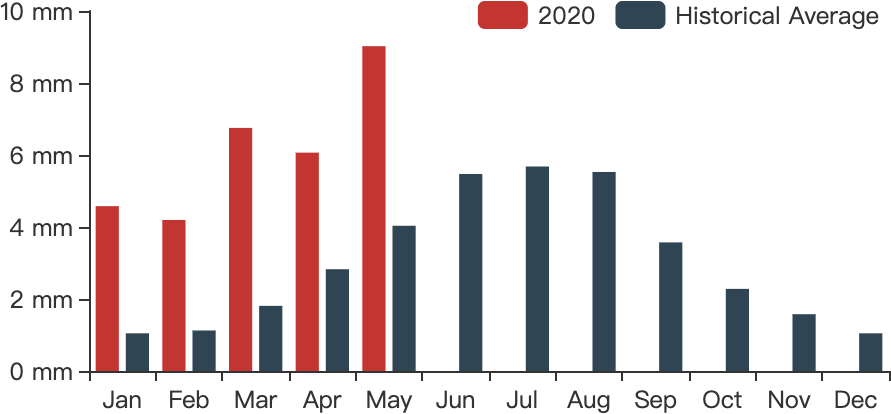}
    \vspace{0.3em}
    \end{minipage}
    }
    
    \caption{Significantly larger precipitation in 2020, in both Wuhan and China.}
    \label{fig:history_precipitation}
\end{figure}

\begin{figure}
    \subfigure[]{ 
    \begin{minipage}[]{0.5\linewidth}
    \flushleft
    \includegraphics[width=0.98\linewidth]{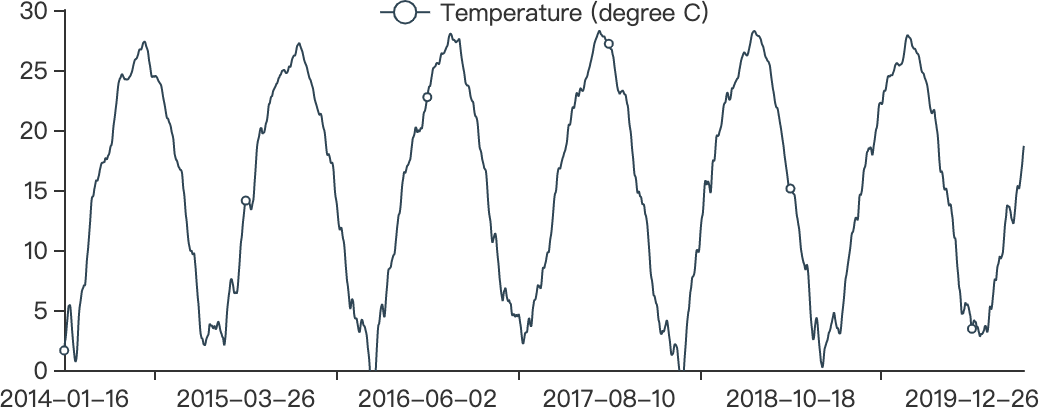}
    \vspace{0.3em}
    \end{minipage}
    }
    \subfigure[]{ 
    \begin{minipage}[]{0.5\linewidth}
    \flushright
    \includegraphics[width=0.98\linewidth]{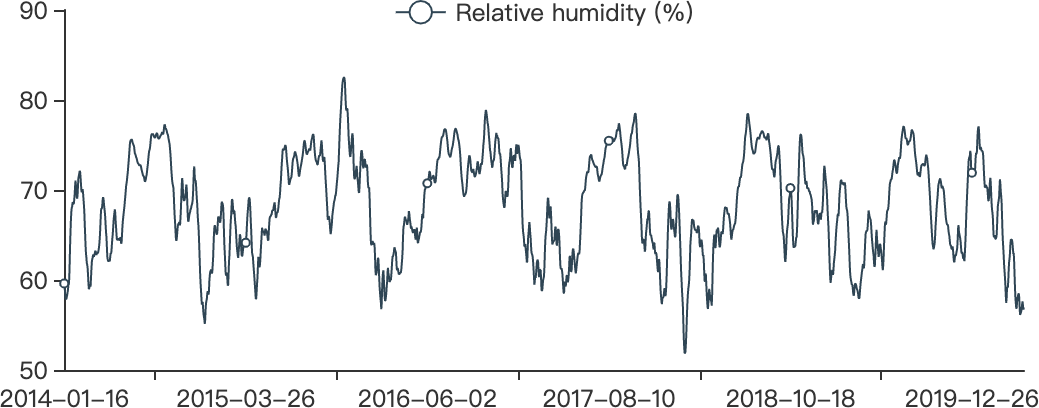}
    \vspace{0.3em}
    \end{minipage}
    }
    \\
    \vspace{0.3em}
    \\
    \subfigure[]{ 
    \begin{minipage}[]{0.5\linewidth}
    \flushleft
    \includegraphics[width=0.98\linewidth]{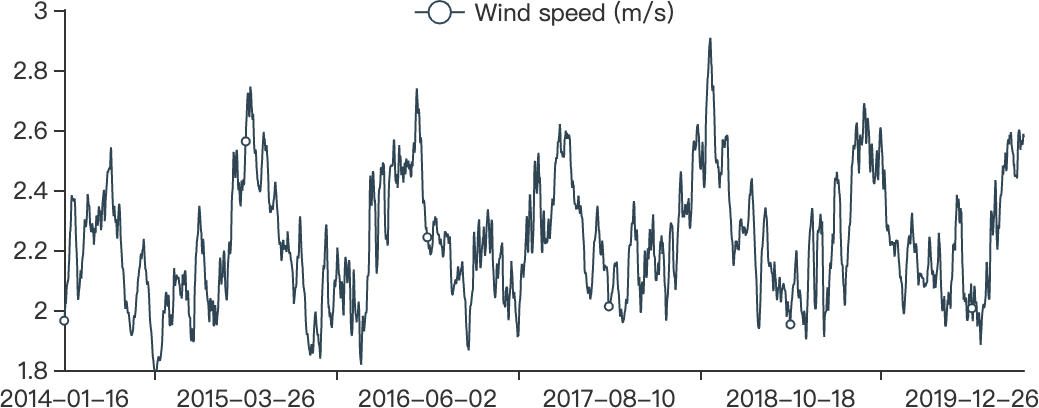}
    \vspace{0.3em}
    \end{minipage}
    }
    \subfigure[]{ 
    \begin{minipage}[]{0.5\linewidth}
    \flushright
    \includegraphics[width=0.98\linewidth]{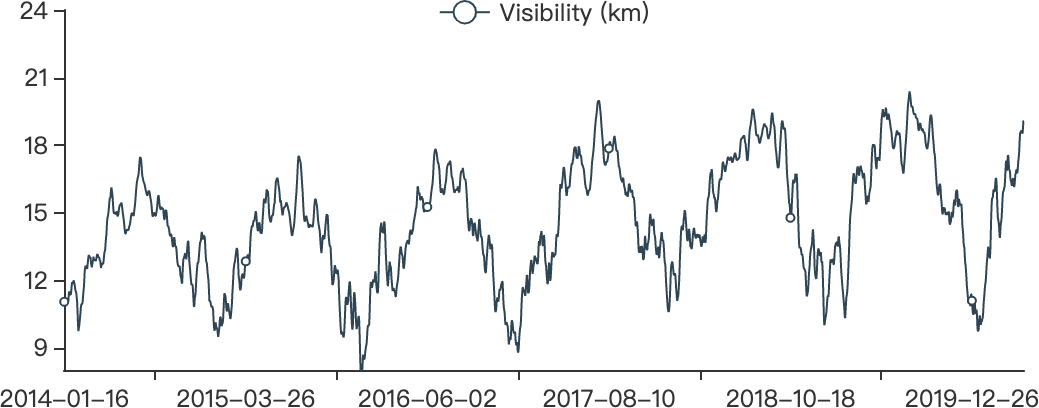}
    \vspace{0.3em}
    \end{minipage}
    }
    
    \caption{Historical weather statistics of China, from 2014 to 2020.}
    \label{fig:history_others}
\end{figure}

\emph{We conduct time-series analysis to show that precipitation plays an important role in the spread of COVID-19 pandemic.}
To better understand the impact of precipitation in 2020, we first investigate the weather statistics in recent years, as shown in Figs. \ref{fig:history_precipitation} and \ref{fig:history_others}. Figs. \ref{fig:history_precipitation}(a) and (b) show that there is abnormally larger precipitation in 2020 in Wuhan and China. Fig. \ref{fig:history_others} reveals that the other weather factors including temperature, relative humidity, wind speed, and visibility are consistent with the historical data. Figs. \ref{fig:history_precipitation}(c) and (d) compare the precipitation of 2020 and the historical average, in Wuhan and China, respectively. The daily precipitation of the first five months in 2020 is 7.7mm(240\%) and 4.0mm(182\%) more than the historical average in Wuhan and China, respectively.

\begin{figure}
    \subfigure[]{ 
    \begin{minipage}[]{0.5\linewidth}
    \flushleft
    \includegraphics[width=0.96\linewidth]{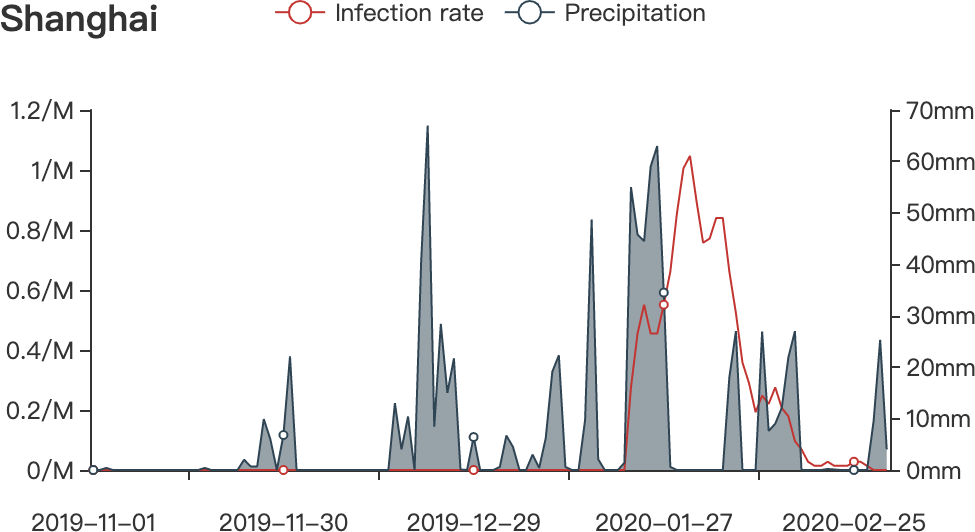}
    \vspace{0.0em}
    \end{minipage}
    }
    \subfigure[]{ 
    \begin{minipage}[]{0.5\linewidth}
    \flushright
    \includegraphics[width=0.96\linewidth]{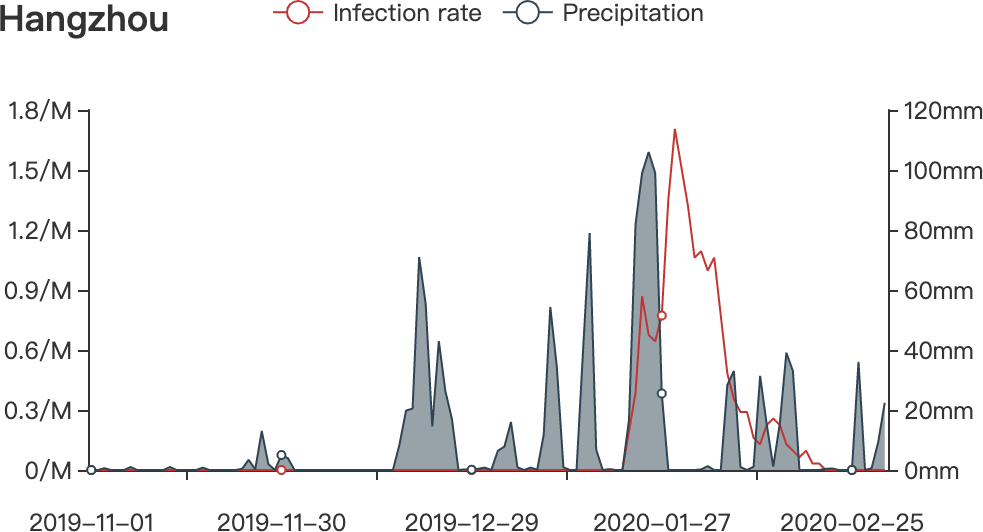}
    \vspace{0.0em}
    \end{minipage}
    }
    \\ \vspace{0.5em} \\
    \subfigure[]{ 
    \begin{minipage}[]{0.5\linewidth}
    \flushleft
    \includegraphics[width=0.96\linewidth]{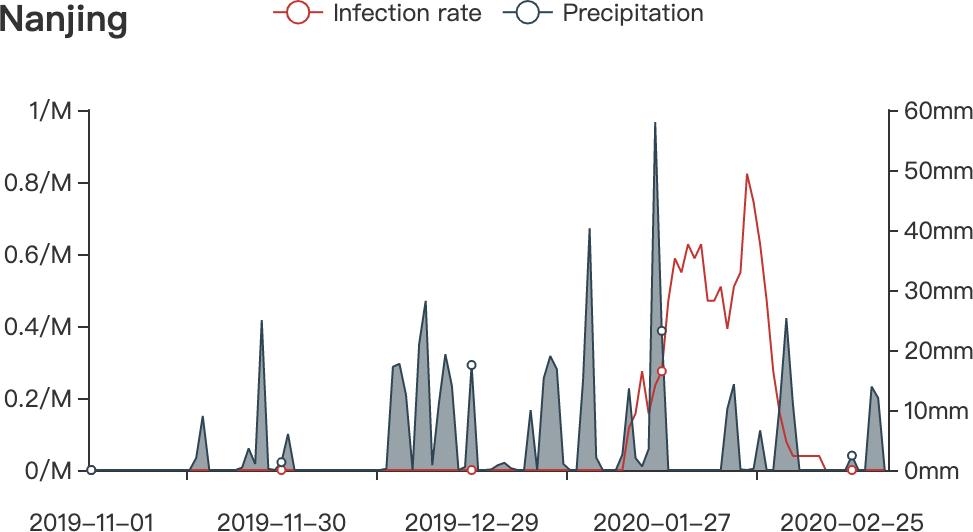}
    \vspace{0.3em}
    \end{minipage}
    }
    \subfigure[]{ 
    \begin{minipage}[]{0.5\linewidth}
    \flushright
    \includegraphics[width=0.96\linewidth]{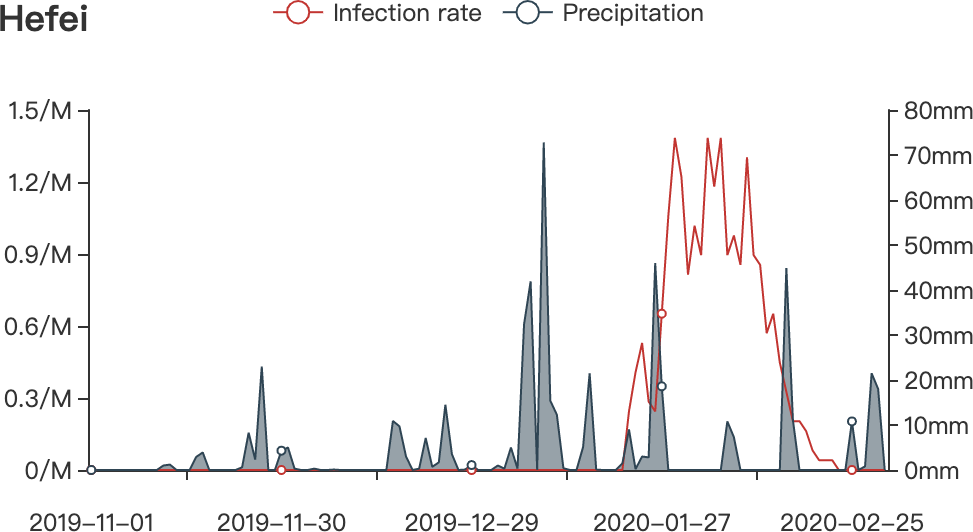}
    \vspace{0.3em}
    \end{minipage}
    }
    \\ \vspace{0.5em} \\
    \subfigure[]{ 
    \begin{minipage}[]{0.5\linewidth}
    \flushleft
    \includegraphics[width=0.96\linewidth]{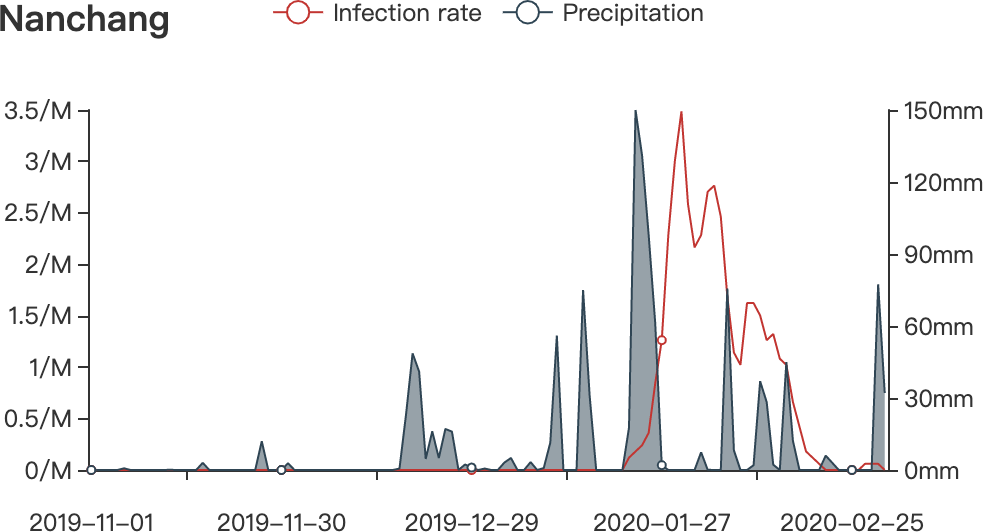}
    \vspace{0.0em}
    \end{minipage}
    }
    \subfigure[]{ 
    \begin{minipage}[]{0.5\linewidth}
    \flushright
    \includegraphics[width=0.96\linewidth]{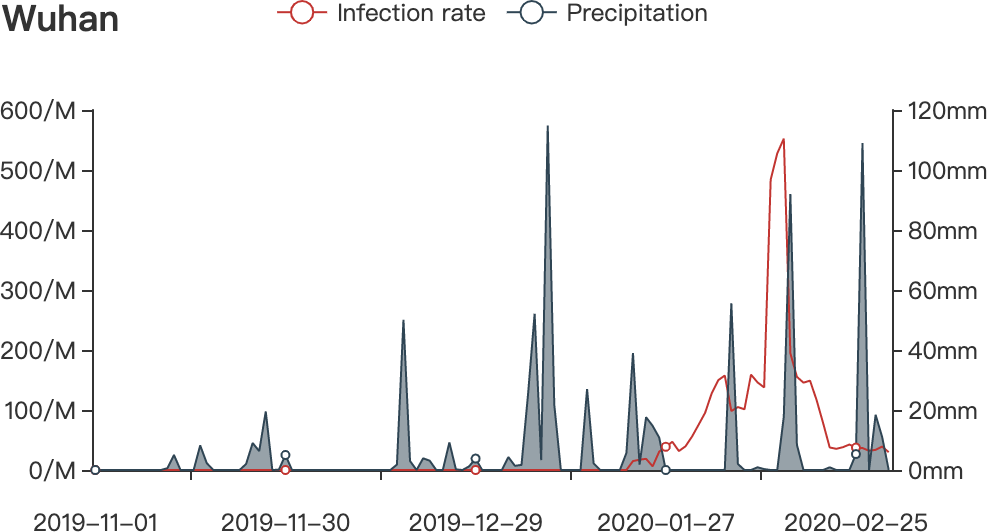}
    \vspace{0.0em}
    \end{minipage}
    }
    \\ \vspace{0.5em} \\
    \subfigure[]{ 
    \begin{minipage}[]{0.5\linewidth}
    \flushleft
    \includegraphics[width=0.96\linewidth]{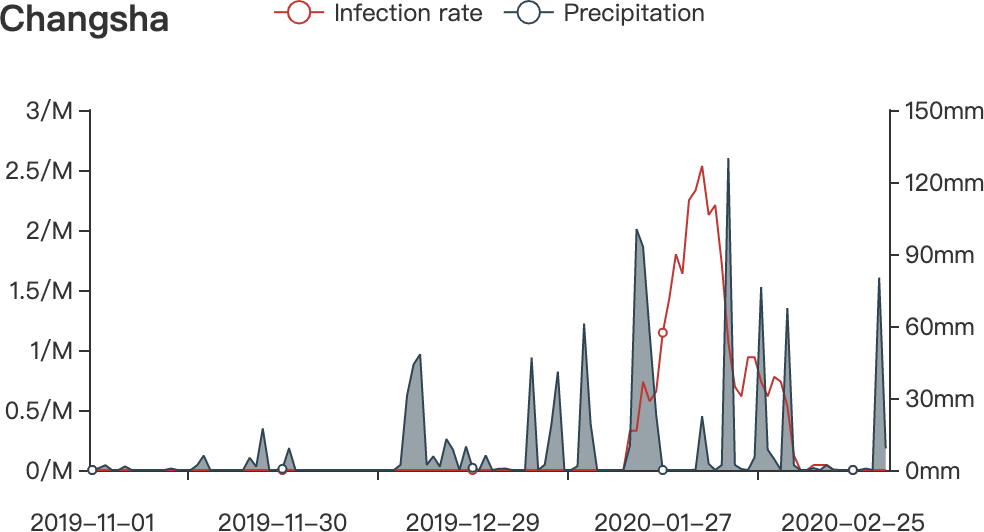}
    \vspace{0.0em}
    \end{minipage}
    }
    \subfigure[]{ 
    \begin{minipage}[]{0.5\linewidth}
    \flushright
    \includegraphics[width=0.96\linewidth]{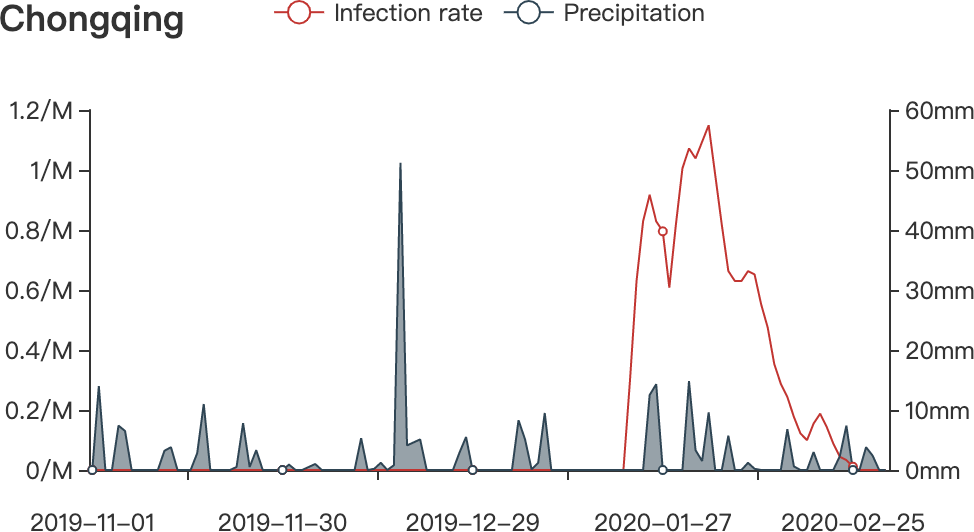}
    \vspace{0.0em}
    \end{minipage}
    }
    \caption{Time-series analysis on the relationship between precipitation and outbreak of COVID-19. Generally, there is significantly larger precipitation before the outbreak of COVID-19 pandemic in eight cities along the Yangtze River, shown from east to west respectively.}
    \label{fig:precipitaton_before_covid19}
\end{figure}

In Fig. \ref{fig:precipitaton_before_covid19}, we study the relationship between precipitation and infection rate in eight cities along the Yangtze River. We find that there is consistently larger precipitation in Shanghai, Hangzhou, Nanjing, Hefei, Nanchang, Wuhan, and Changsha, before the outbreak of COVID-19 pandemic. In Wuhan, the precipitation becomes much larger starting from the January of 2020. Fig. \ref{fig:precipitaton_before_covid19} to some extent validates our assumption that precipitation plays an important role in the outbreak of COVID-19 pandemic.

\subsection{Weather-based Prediction of the Second Outbreak}
\label{sec:result_prediction}

\begin{figure}
    \centering
    \includegraphics[width=0.5\linewidth]{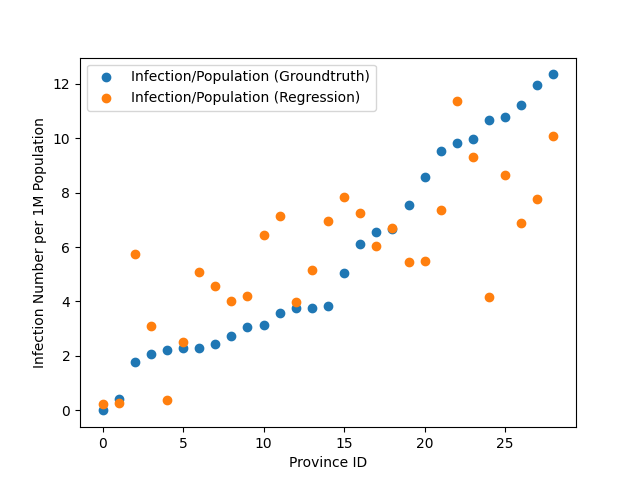}
    \caption{Results of our outbreak prediction model, using multivariate regression method to transform the weather factors to infection rate.}
    \label{fig:multivariate}
\end{figure}

\emph{We build weather-based prediction model to estimate the infection rates of different provinces in Mainland China if there is a second outbreak of coronavirus.}
Using the multivariate regression method discussed in Section \ref{sec:multivariate}, Fig. \ref{fig:multivariate} shows the regression results of the second outbreak of Chinese provinces. Each data point denotes a province, and the y-axis shows the infection rate. We sort the data points according to the real values of the infection rate. The regression values show a similar trend with the real values, showing that the regression model based on weather statistics can modestly estimate the trend of COVID-19 outbreak. The containment policies of different cities in Mainland China is generally consistent, such that we do not include the social factors in this predictive model.

\begin{figure}
    \subfigure[January]{ 
    \begin{minipage}[]{0.5\linewidth}
    \flushleft
    \includegraphics[width=0.96\linewidth]{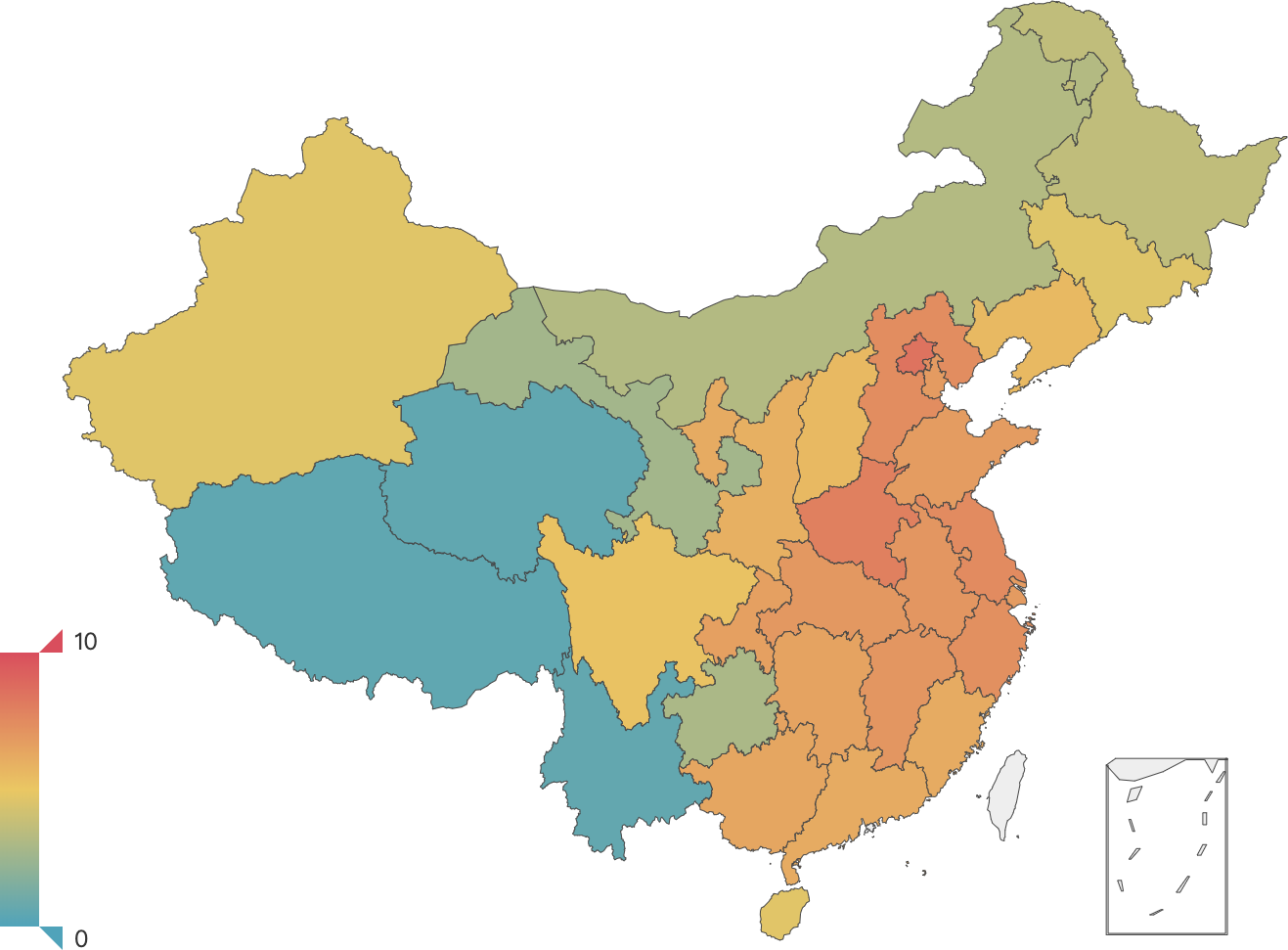}
    \vspace{0em}
    \end{minipage}
    }
    \subfigure[March]{ 
    \begin{minipage}[]{0.5\linewidth}
    \flushright
    \includegraphics[width=0.96\linewidth]{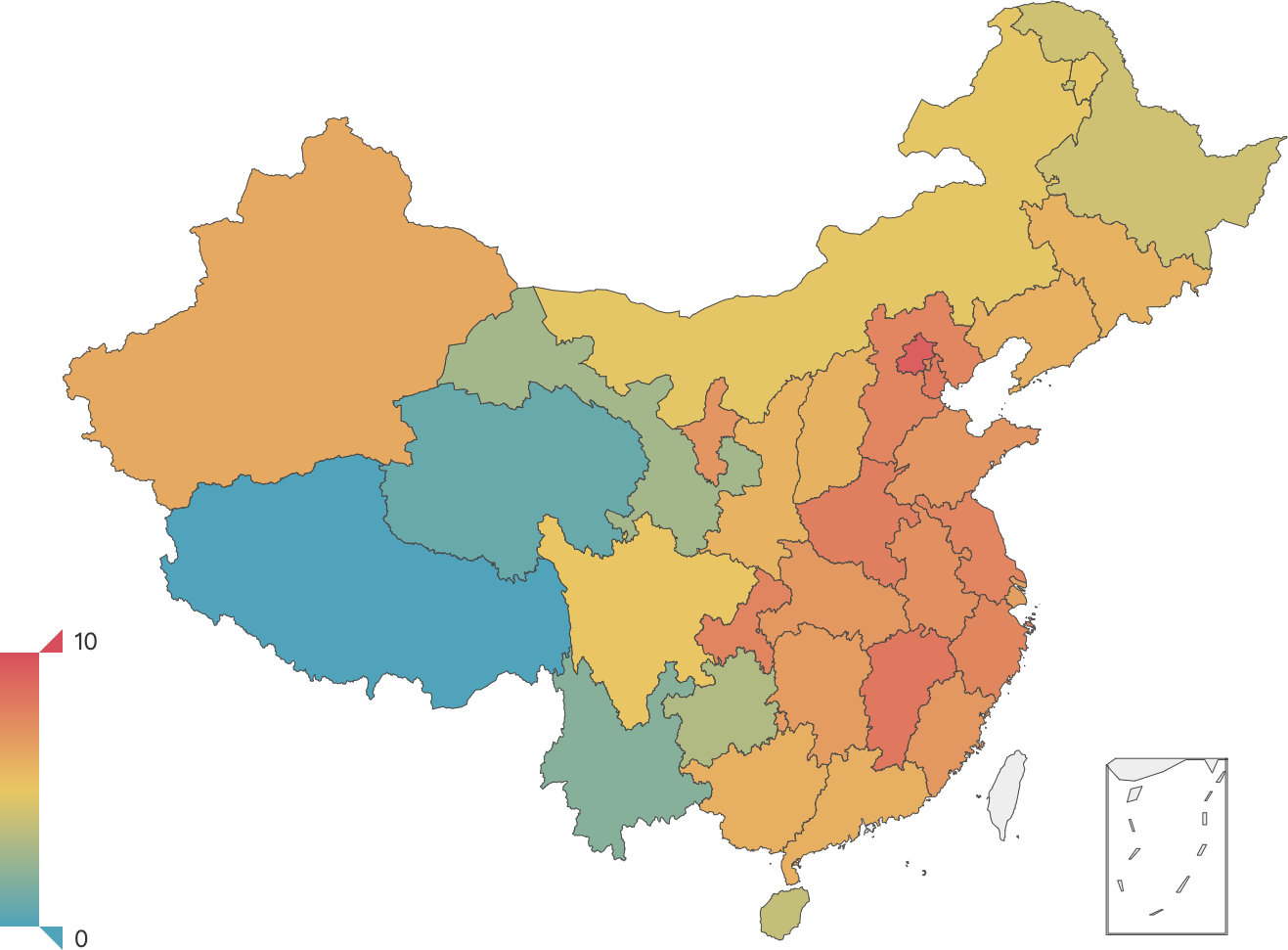}
    \vspace{0em}
    \end{minipage}
    }
    \\
    \vspace{0.5em}
    \\
    \subfigure[May]{ 
    \begin{minipage}[]{0.5\linewidth}
    \flushleft
    \includegraphics[width=0.96\linewidth]{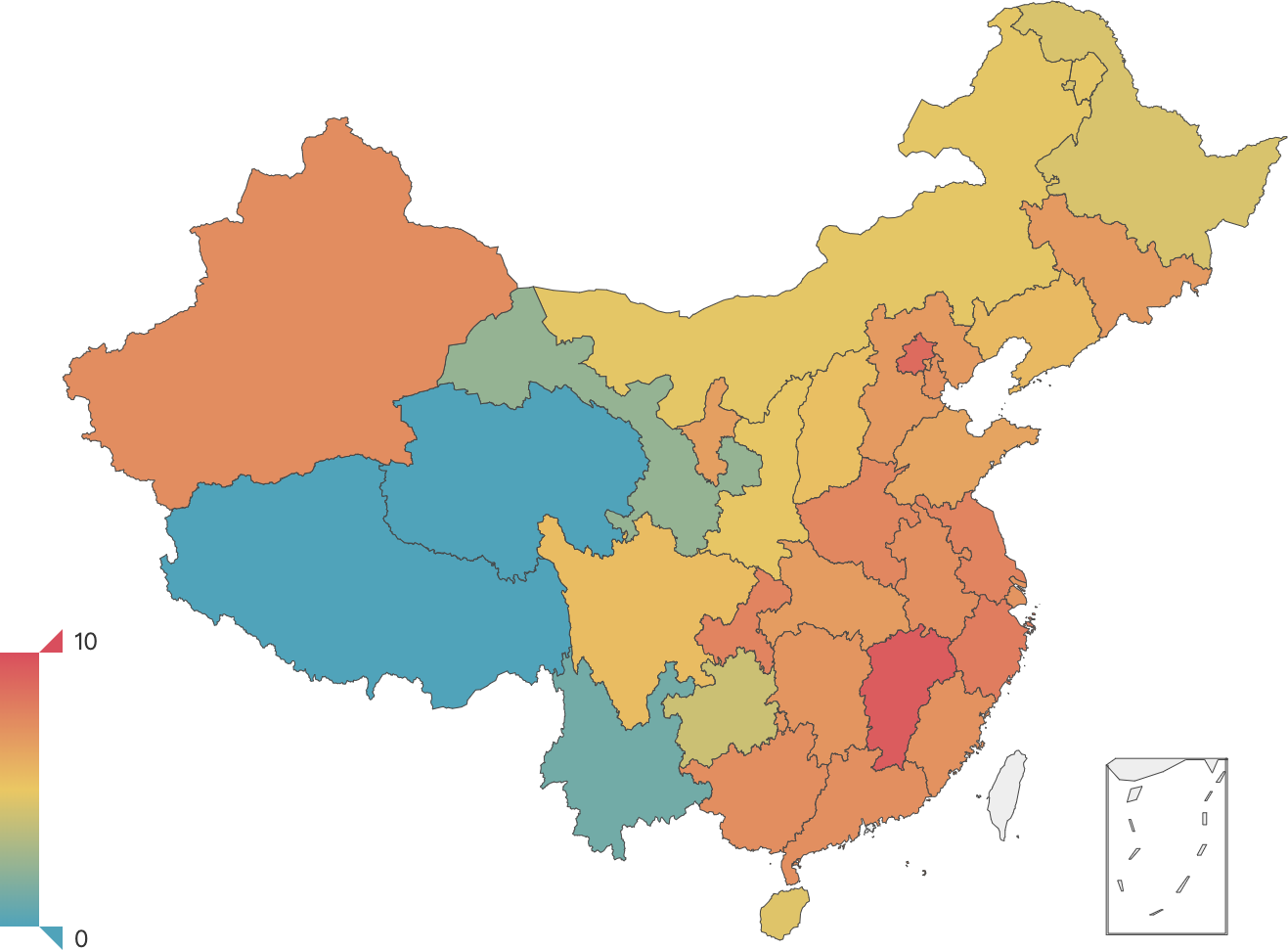}
    \vspace{0em}
    \end{minipage}
    }
    \subfigure[July]{ 
    \begin{minipage}[]{0.5\linewidth}
    \flushright
    \includegraphics[width=0.96\linewidth]{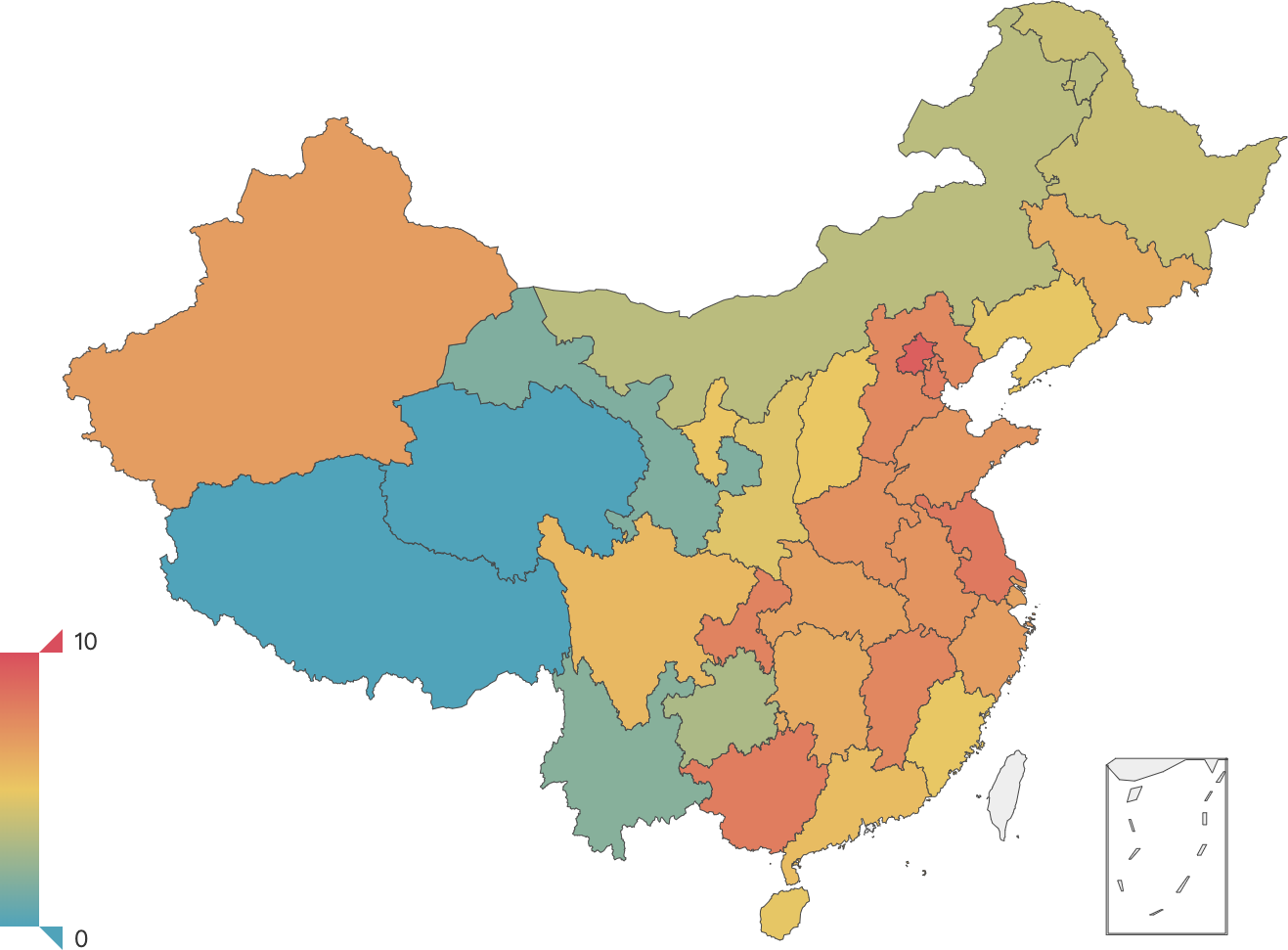}
    \vspace{0em}
    \end{minipage}
    }
    \\
    \vspace{0.5em}
    \\
    \subfigure[September]{ 
    \begin{minipage}[]{0.5\linewidth}
    \flushleft
    \includegraphics[width=0.96\linewidth]{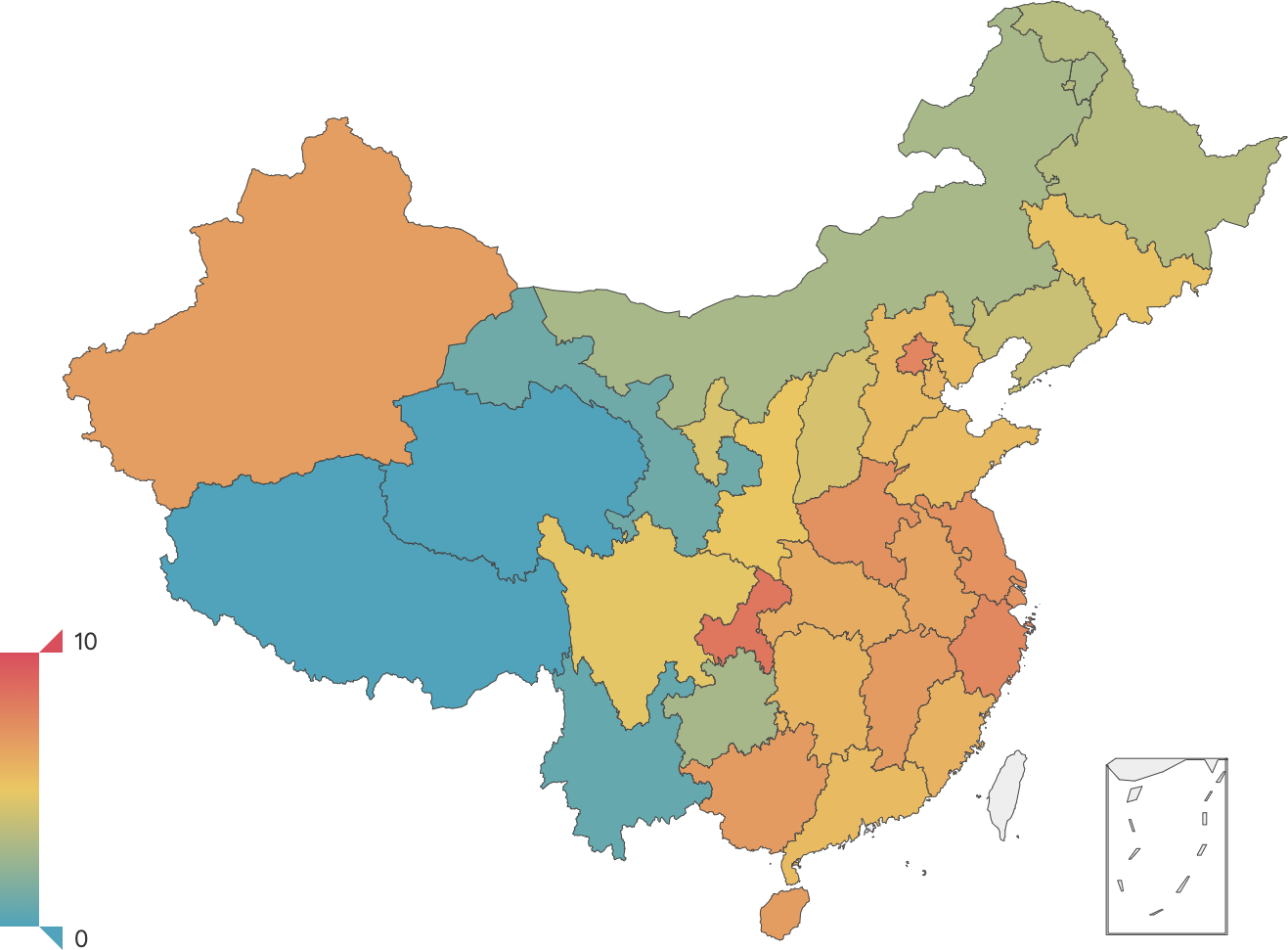}
    \vspace{0em}
    \end{minipage}
    }
    \subfigure[November]{ 
    \begin{minipage}[]{0.5\linewidth}
    \flushright
    \includegraphics[width=0.96\linewidth]{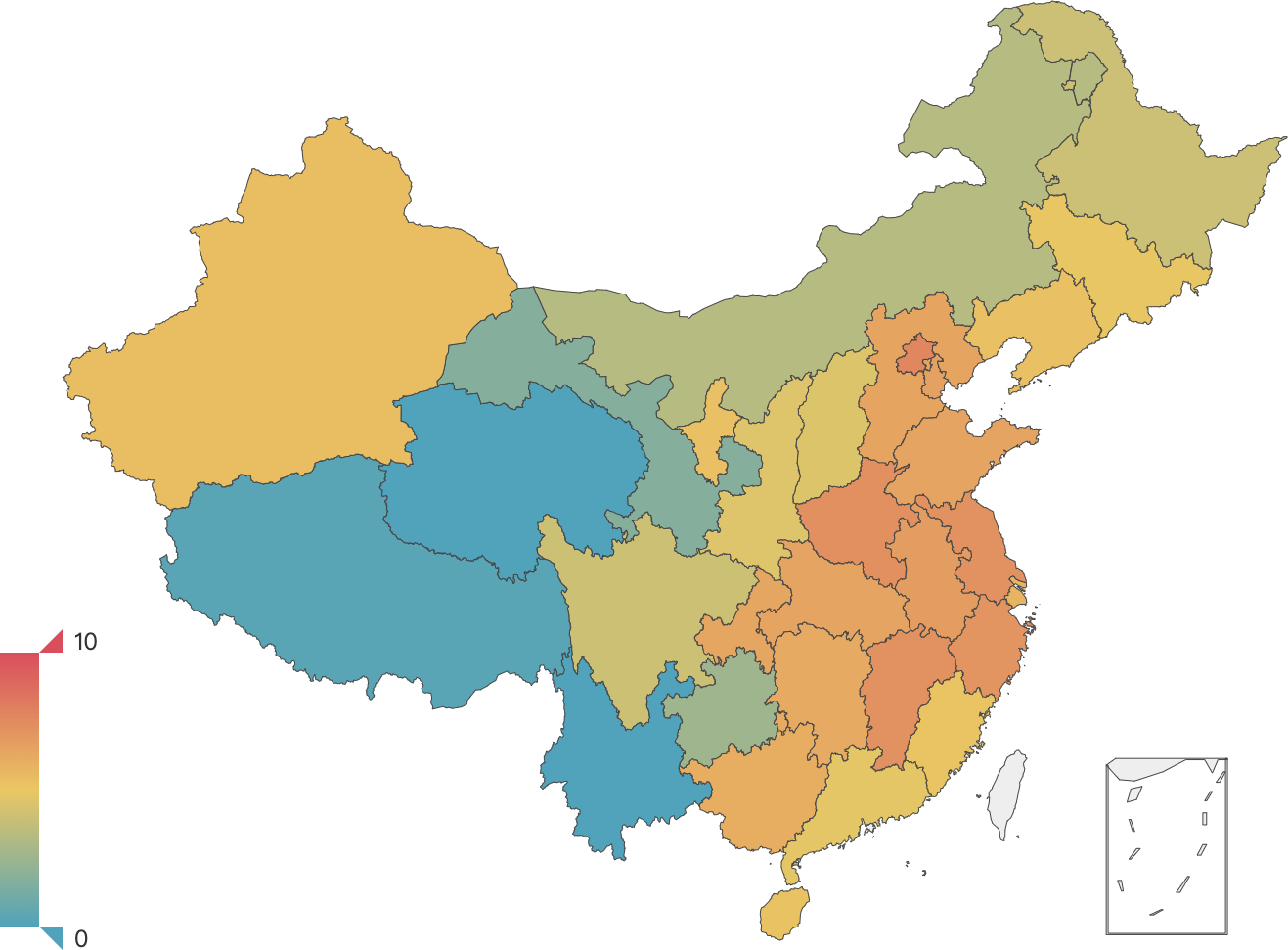}
    \vspace{0em}
    \end{minipage}
    }
    
    \caption{The estimated infection numbers per 1 million people with respect to different months, if there is a second coronavirus outbreak in Mainland China. }
    \label{fig:secondoutbreak_map}
\end{figure}

\begin{figure}
    \includegraphics[width=1\linewidth]{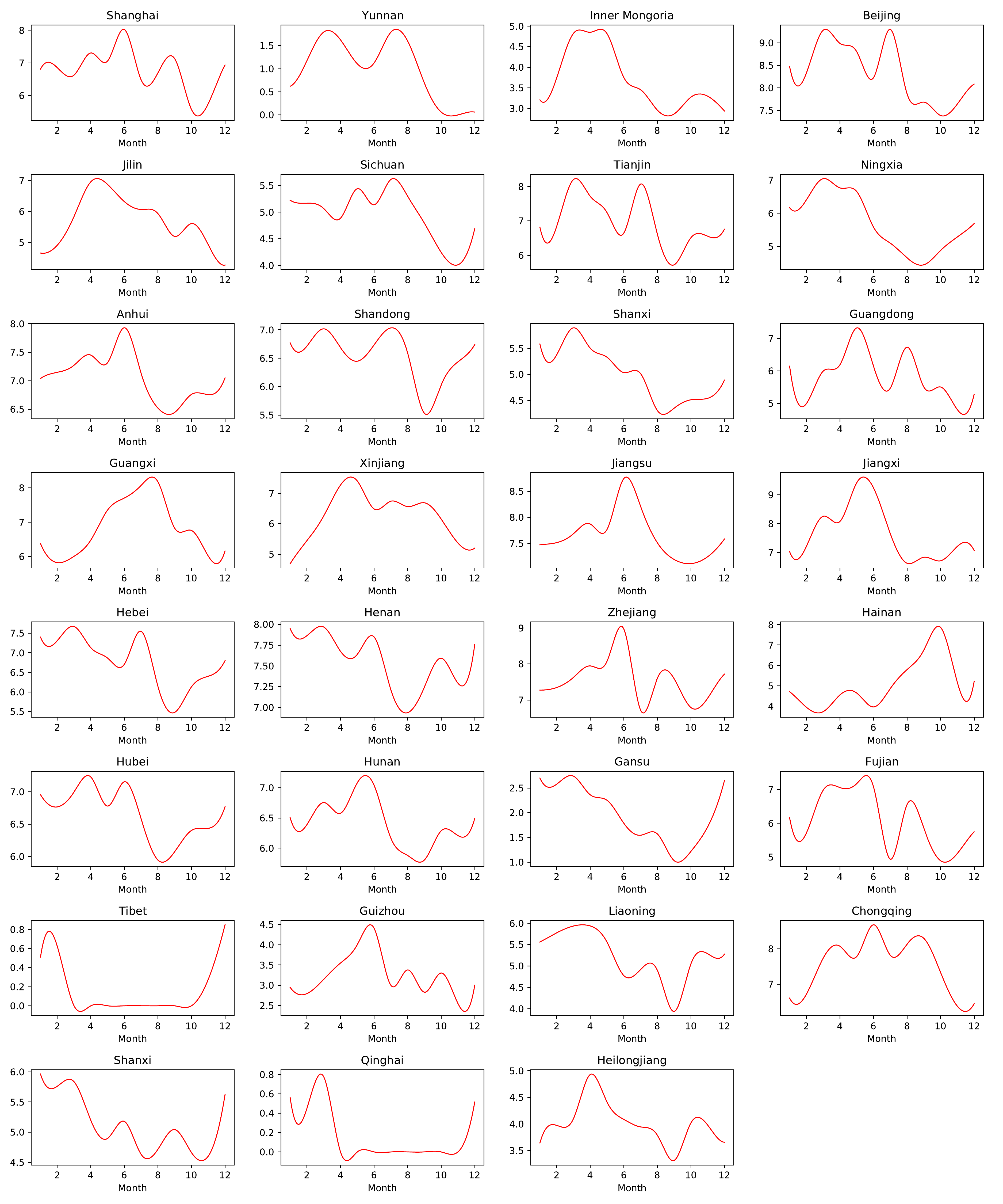}
    \caption{The estimated infection numbers per 1 million people with respect to different provinces, if there is a second coronavirus outbreak in Mainland China. }
    \label{fig:secondoutbreak_temporal}
\end{figure}

\emph{The outbreak prediction model shows that the danger degrees and the dangerous periods of coronavirus pandemic are diverse with respect to different geographical areas.}
Fig. \ref{fig:secondoutbreak_map} shows the heat maps of estimated local infection rate in different months. In Fig. \ref{fig:secondoutbreak_map}, there are not distinct differences among estimated infection rates of different months. Generally, the central, eastern, and southern provinces of China have larger infection rates compared with the other areas. Fig. \ref{fig:secondoutbreak_temporal} shows more detailed results on all individual provinces. We observe that different provinces have various temporal trends. For instance the most dangerous months of Beijing are March and July, while the most dangerous month of Hainan Province is October. The peak values of different provinces also vary largely. The peak infection rate of Beijing is about $9.5$ (the estimated infection numbers per 1 million people), while the peak infection rate of Yunnan Province is only about $2.0$. The weather-based estimation result suggests that the degree and temporal window of danger of COVID-19 are quite diverse in different geographical areas.

\section{Conclusion}

In this paper, we study the relationships between six weather factors (temperature, relative humidity, precipitation, wind speed, air pressure, visibility, inflow) and the infection statistics of COVID-19 pandemic in 250 cities of Mainland China. The correlation analysis and partial correlation analysis suggest that various weather factors are related with the infection rate, where precipitation is the most significant weather factor in the spread of COVID-19 in China. Further, the time-series analysis shows that precipitation plays an important role in the spread of COVID-19 pandemic in Mainland China. 

We additionally build a weather-based prediction model to estimate the infection rates of different geographical areas in Mainland China if there is a second outbreak of coronavirus. The prediction model shows that both the danger degrees and the dangerous periods of coronavirus pandemic are diverse with respect to different geographical areas.

\section{Acknowledgement}
The authors thank for the beneficial discussions and comments from Dr. Huaiyu Tian. To protect Baidu users' data privacy, all experiments in this paper were carried out using anonymous data and secure data analytics provided by Baidu Data Federation Platform (Baidu FedCube). For data accesses and usages, please
contact us via \{fedcube, shubang\}@baidu.com.

\bibliography{main}
\bibliographystyle{Science}

\newpage

\begin{table*}[t]
\caption{Partial Correlation Analysis on 250 Chinese Cities}
\label{tab:partial}
\begin{center}
\begin{tabular}{c c c | c c}\hline
$x$ & $y$ & Controlling variable & Coeff. (R) & $p$-value  \\\hline \hline
Inflow & Infection & - &  $0.79$  & $6.4\times10^{-51}$  \\ 
Inflow & Infection rate & - &  $0.39$  & $4.2\times10^{-10}$  \\ 
Inflow & Infection rate & All 6 weather factors &  $0.35$  & $4.8\times10^{-8}$  \\ \hline
Temperature & Infection rate & Inflow &  $0.05$  & $4.3\times10^{-1}$  \\
Relative humidity & Infection rate & Inflow &  $0.13$  & $4.2\times10^{-2}$  \\
Precipitation & Infection rate & Inflow &  $0.21$  & $\mathbf{1.6\times10^{-3}}$  \\
Wind speed & Infection rate & Inflow &  $0.10$  & $1.2\times10^{-1}$  \\
Air pressure & Infection rate & Inflow &  $0.14$  & $3.3\times10^{-2}$  \\
Visibility & Infection rate & Inflow &  $-0.15$  & $2.4\times10^{-2}$  \\
\hline
\end{tabular}
\end{center}
\end{table*}

\begin{figure}
    \subfigure[Temperature - Humidity]{ 
    \begin{minipage}[]{0.31\linewidth}
    \includegraphics[width=0.98\linewidth]{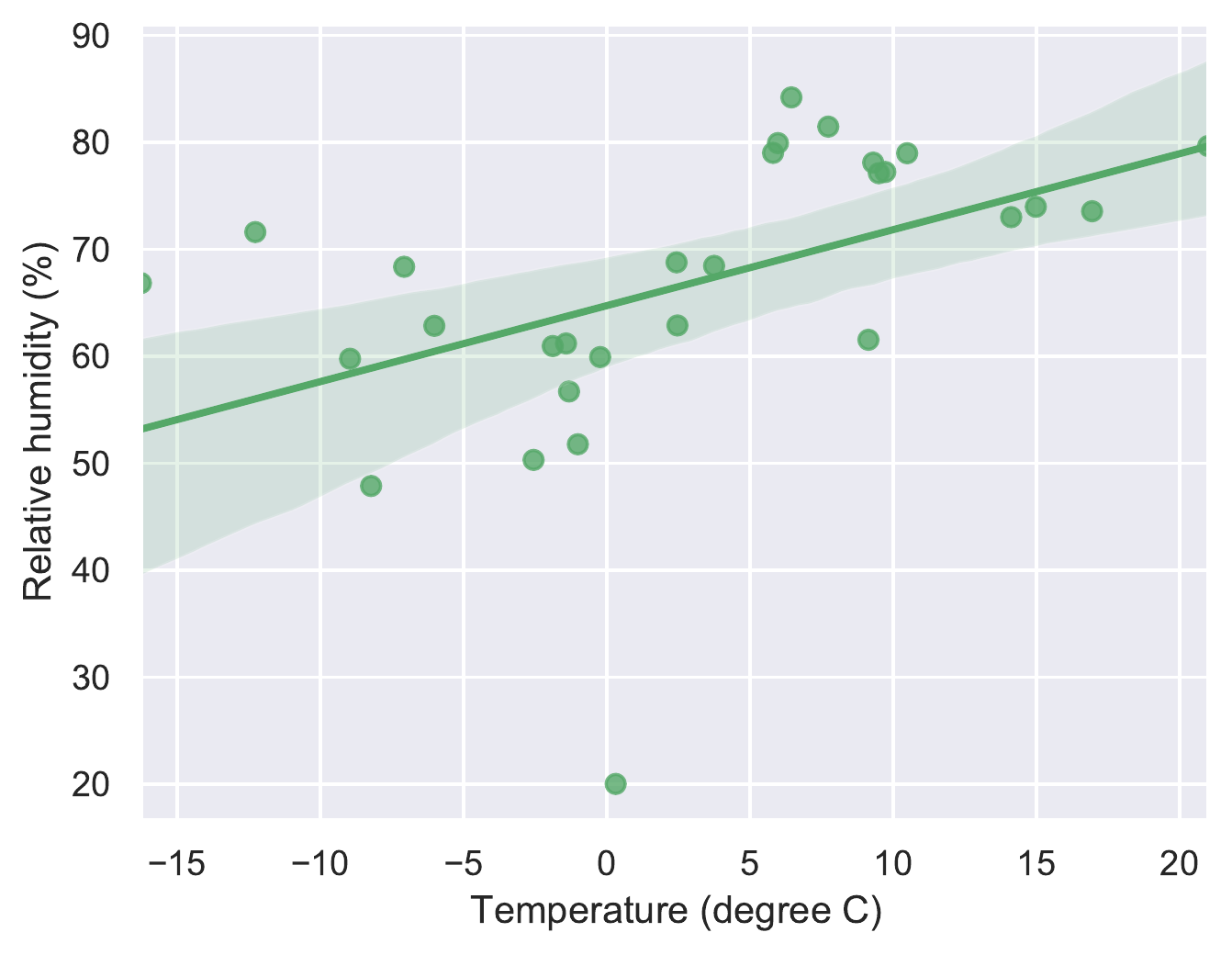}
    \vspace{0.5em}
    \end{minipage}
    }
    \subfigure[Temperature - Precipitation]{ 
    \begin{minipage}[]{0.31\linewidth}
    \includegraphics[width=0.98\linewidth]{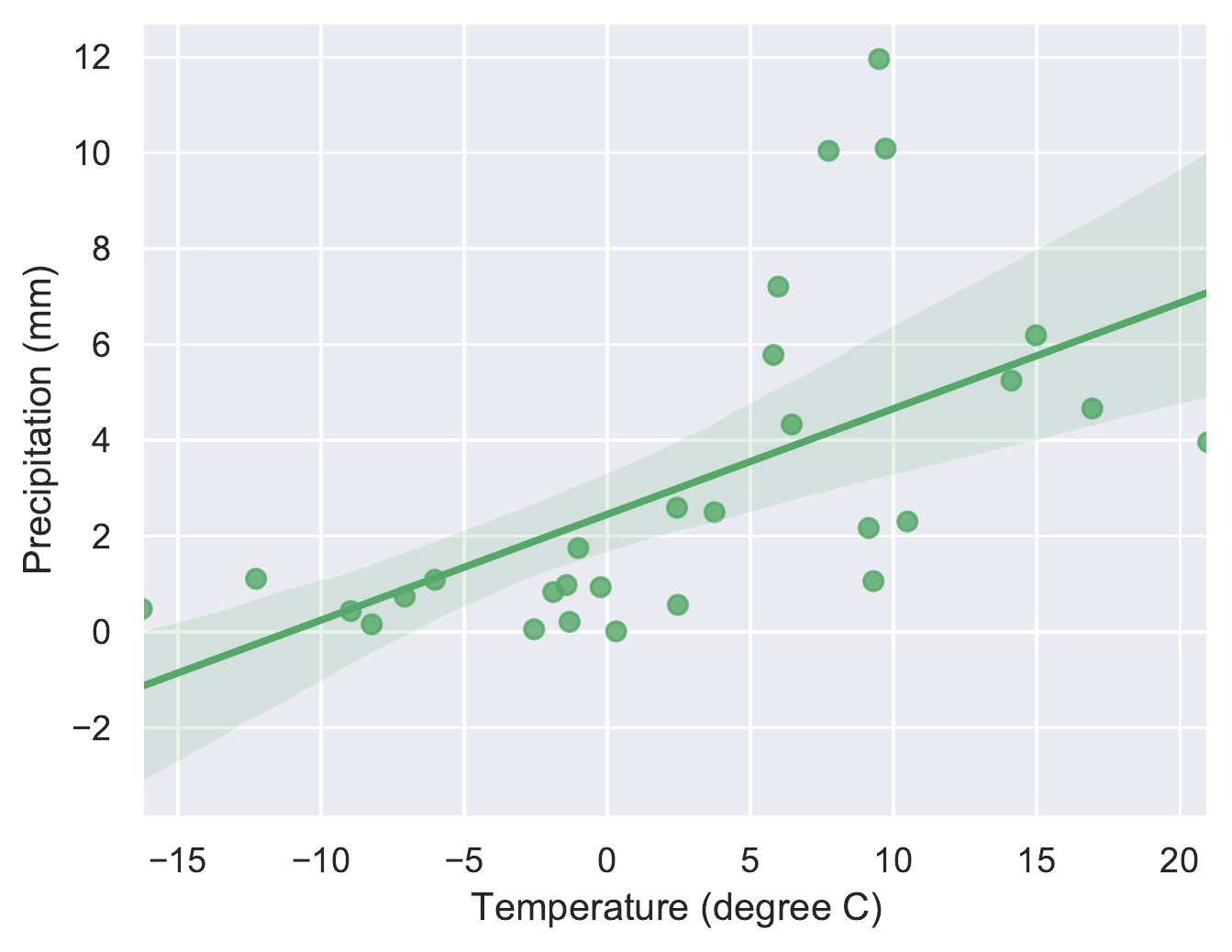}
    \vspace{0.5em}
    \end{minipage}
    }
    \subfigure[Temperature - Air pressure]{ 
    \begin{minipage}[]{0.31\linewidth}
    \includegraphics[width=0.98\linewidth]{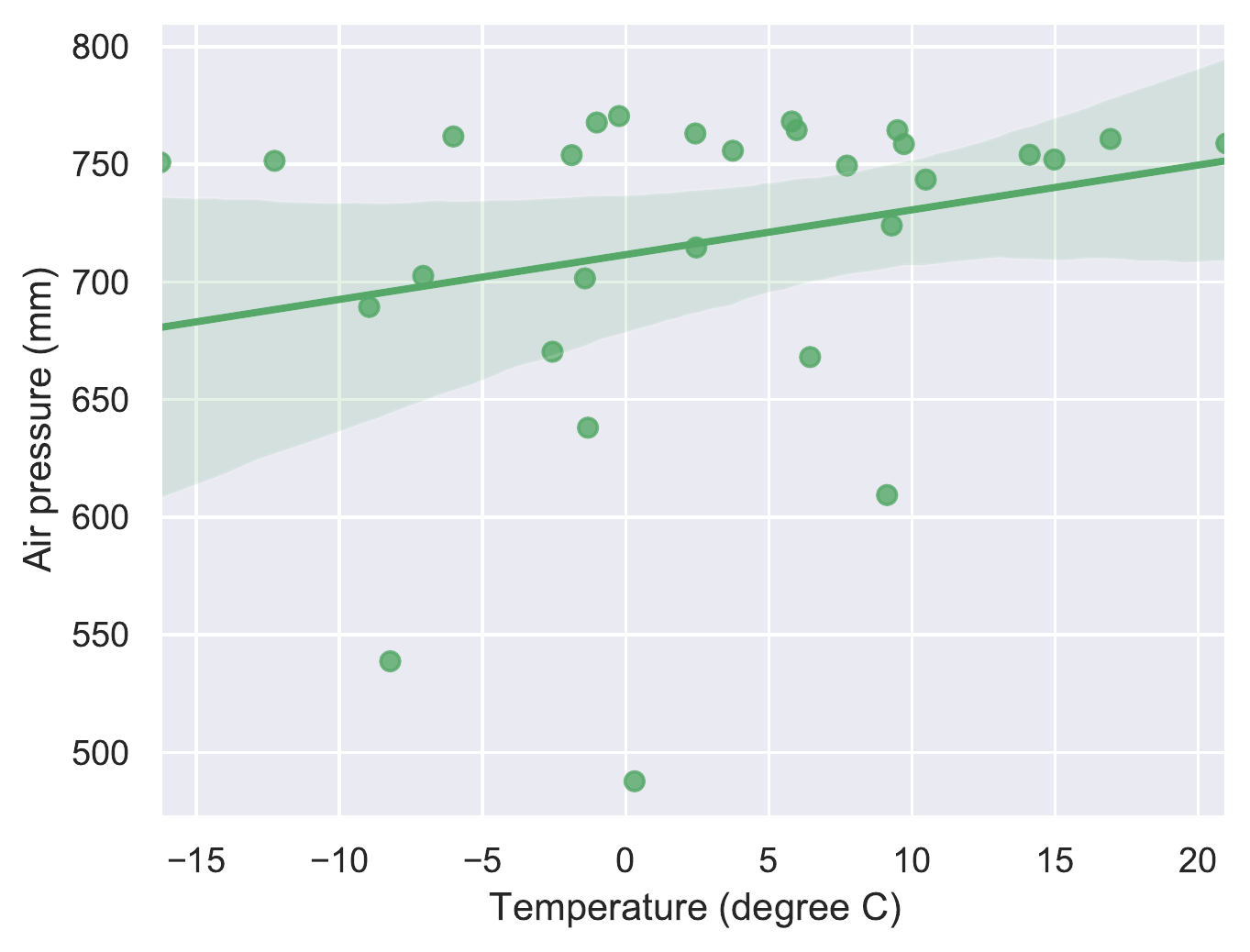}
    \vspace{0.5em}
    \end{minipage}
    }
    \\
    \vspace{0.3em}
    \\
    \subfigure[Humidity - Precipitation]{ 
    \begin{minipage}[]{0.31\linewidth}
    \includegraphics[width=0.98\linewidth]{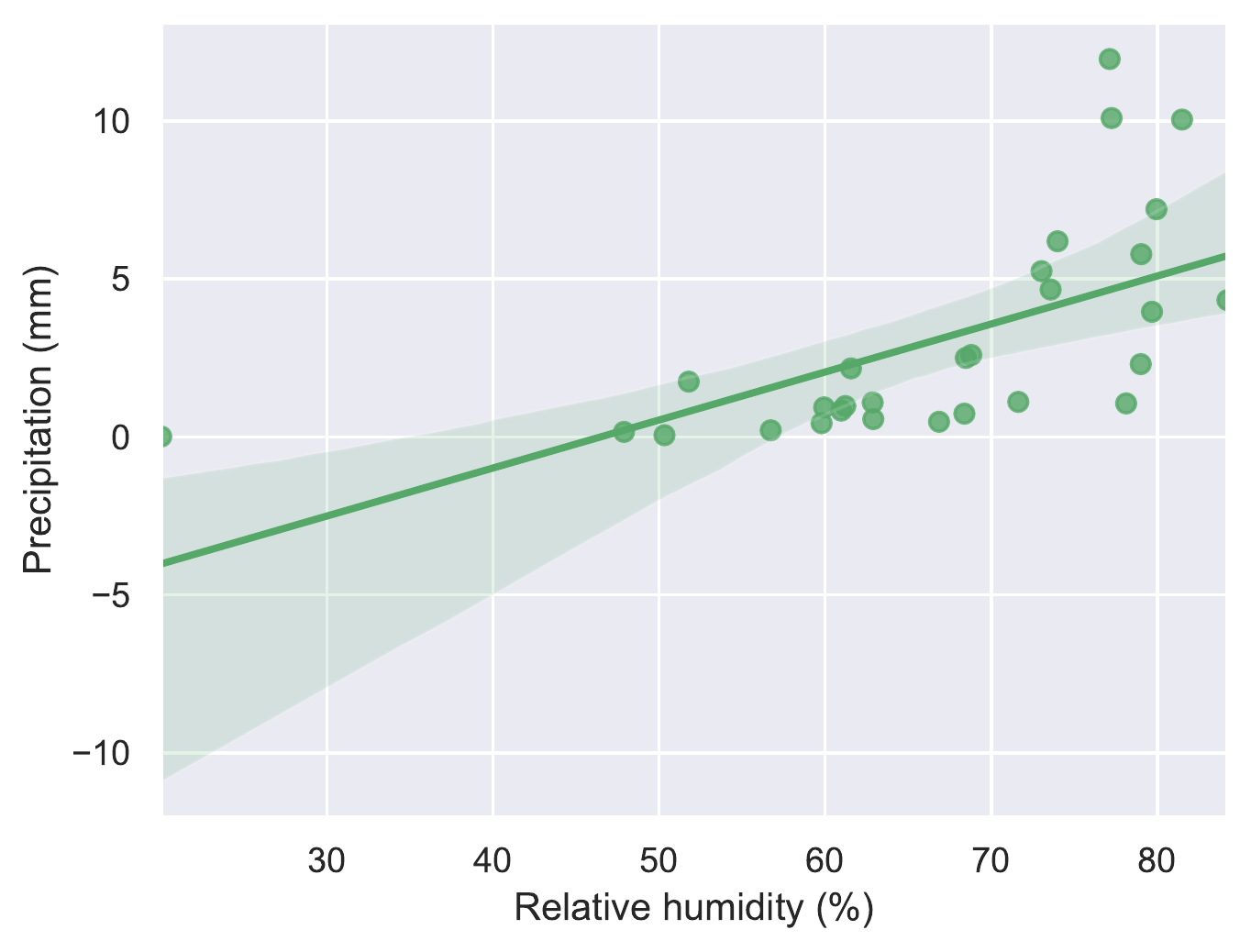}
    \vspace{0.5em}
    \end{minipage}
    }
    \subfigure[Humidity - Air pressure]{ 
    \begin{minipage}[]{0.31\linewidth}
    \includegraphics[width=0.98\linewidth]{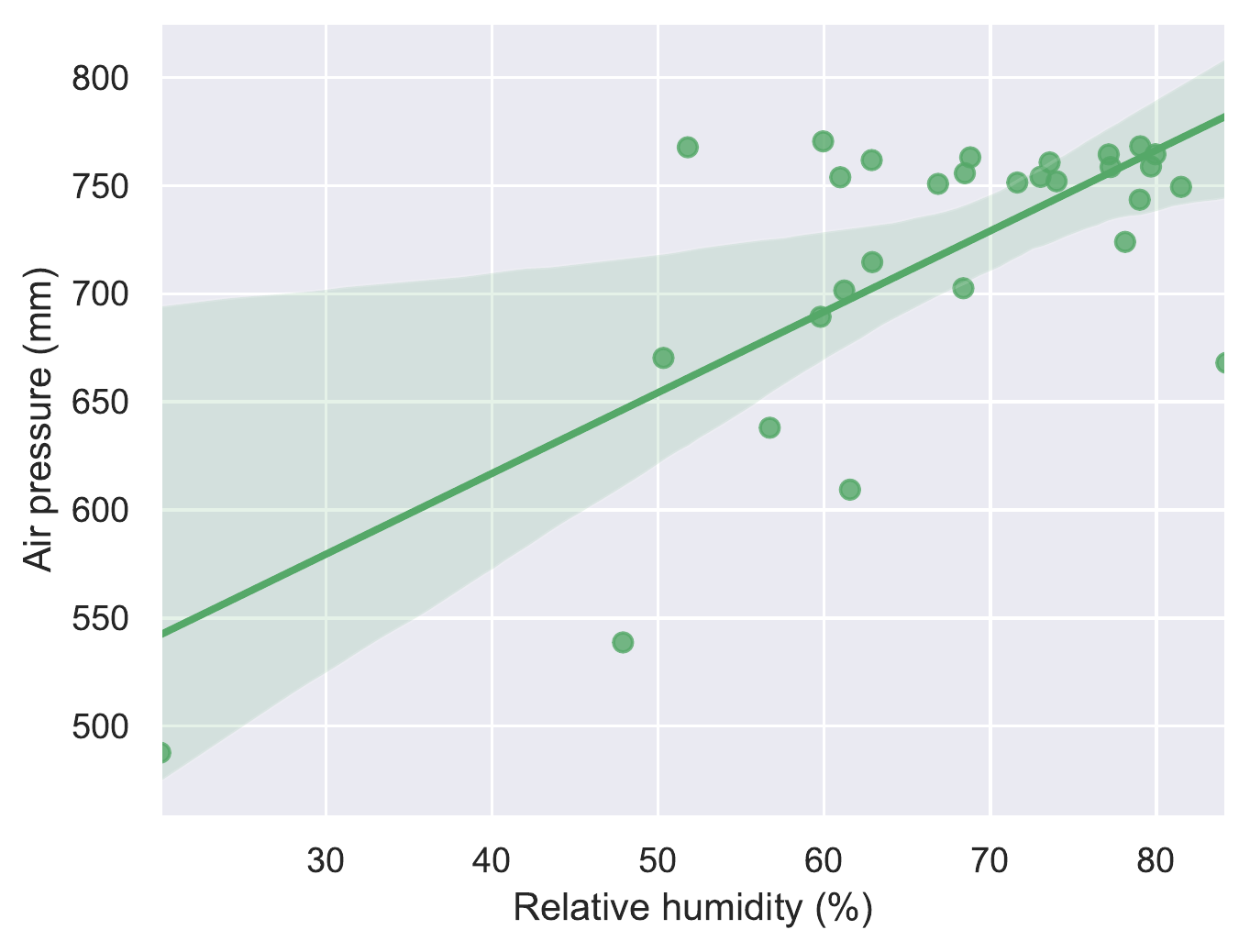}
    \vspace{0.5em}
    \end{minipage}
    }
    \subfigure[Humidity - Visibility]{ 
    \begin{minipage}[]{0.31\linewidth}
    \includegraphics[width=0.98\linewidth]{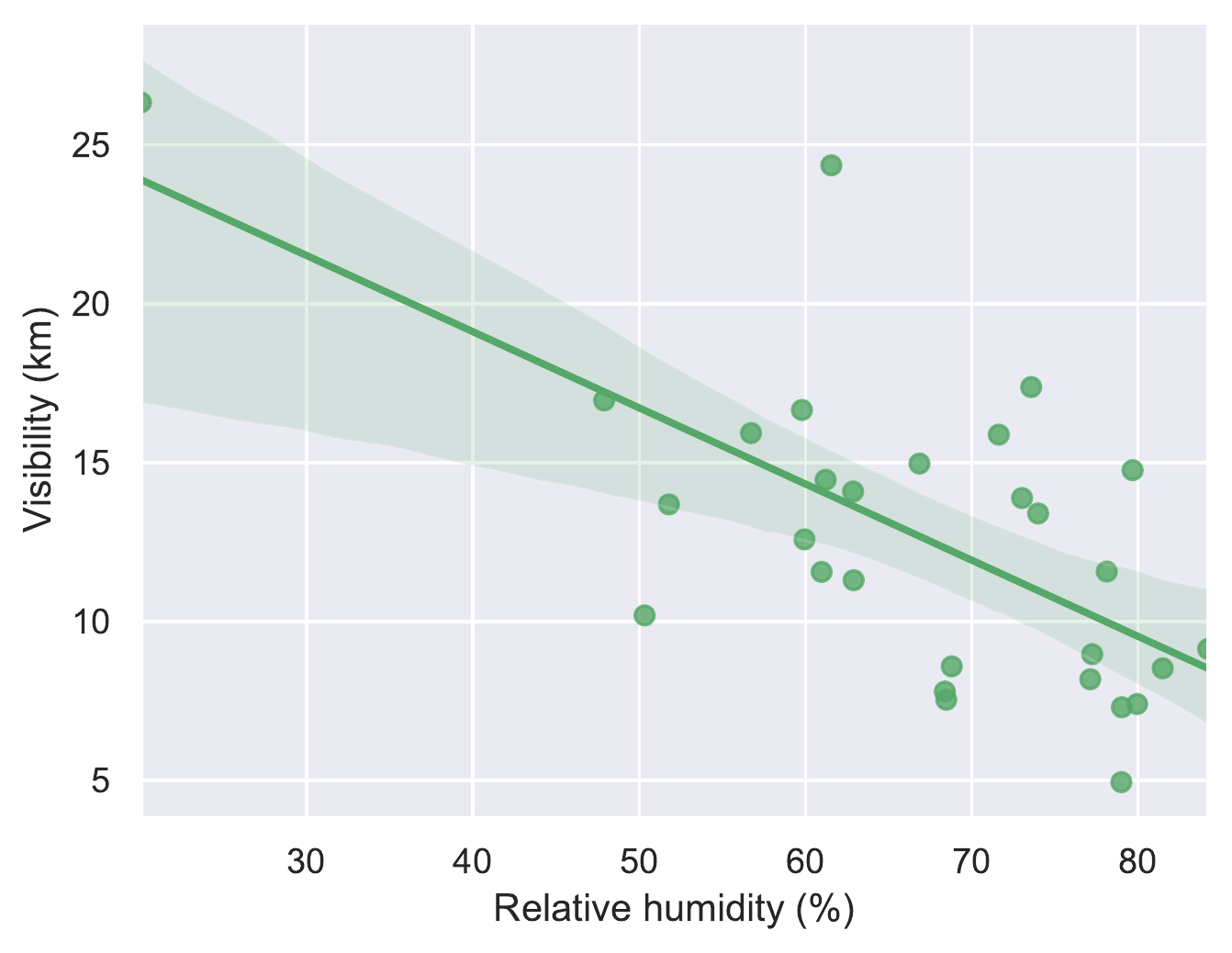}
    \vspace{0.5em}
    \end{minipage}
    }
    \\
    \vspace{0.3em}
    \\
    \subfigure[Precipitation - Air pressure]{ 
    \begin{minipage}[]{0.31\linewidth}
    \includegraphics[width=0.98\linewidth]{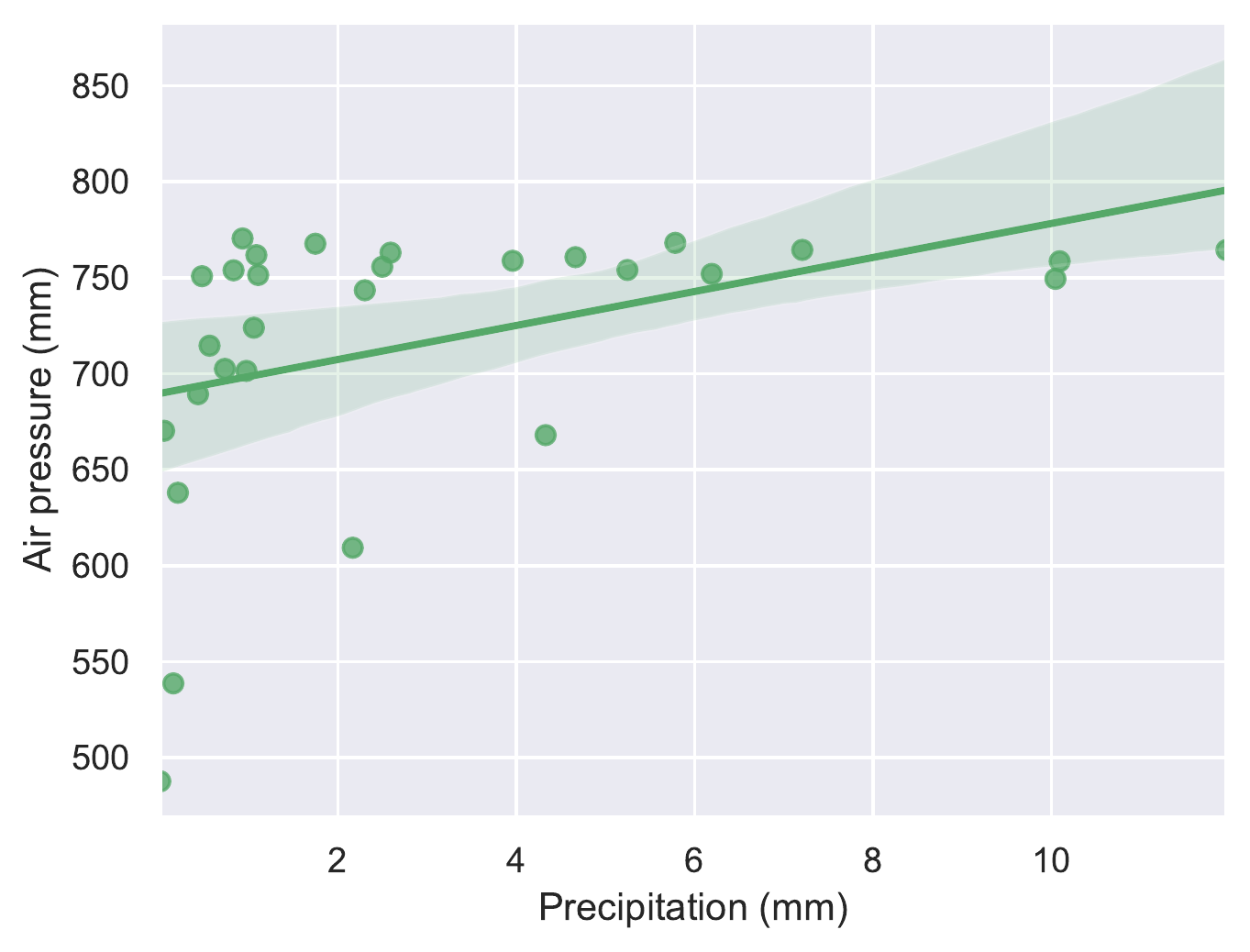}
    \vspace{0.5em}
    \end{minipage}
    }
    \subfigure[Precipitation - Visibility]{ 
    \begin{minipage}[]{0.31\linewidth}
    \includegraphics[width=0.98\linewidth]{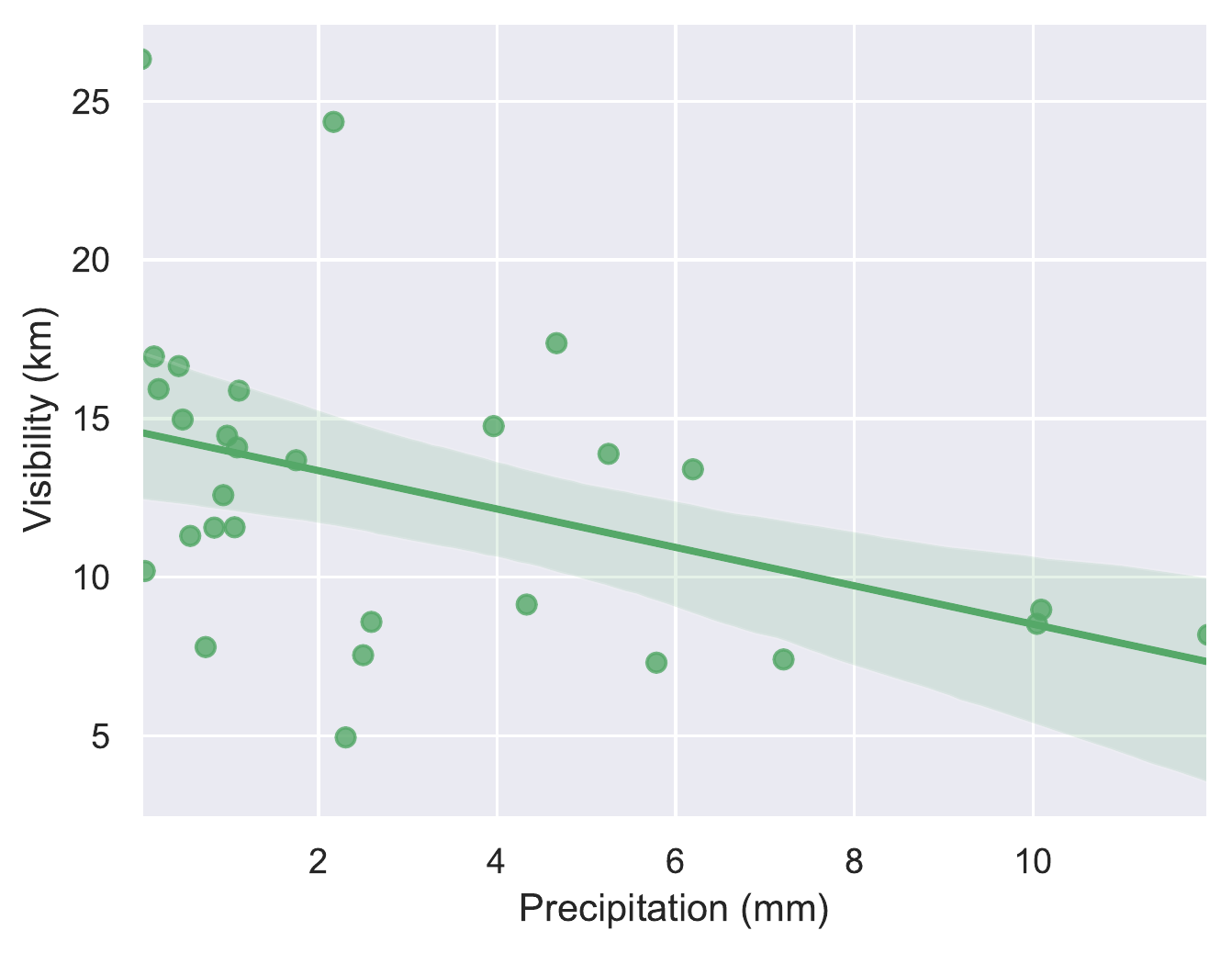}
    \vspace{0.5em}
    \end{minipage}
    }
    \subfigure[Air pressure - Visibility]{ 
    \begin{minipage}[]{0.31\linewidth}
    \includegraphics[width=0.98\linewidth]{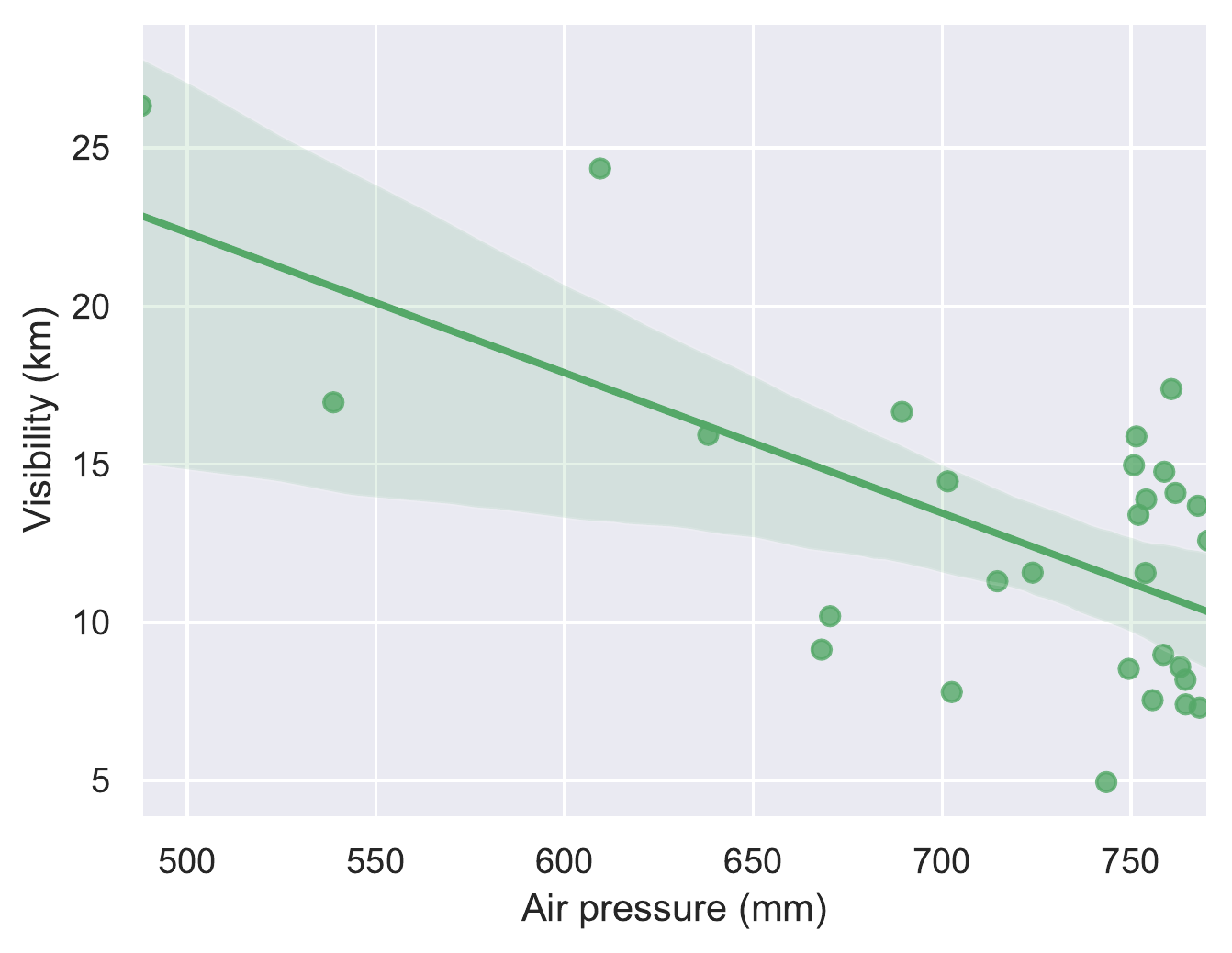}
    \vspace{0.5em}
    \end{minipage}
    }
    \caption{Linear regression results on pairs of weather factors. Only significant results are shown.}
    \label{fig:confidence_weather}
\end{figure}

\end{document}